\documentclass[pdflatex,sn-mathphys-num]{sn-jnl}
\usepackage{graphicx}%
\usepackage{multirow}%
\usepackage{caption}
\usepackage{amsmath,amssymb,amsfonts}%
\usepackage{amsthm}%
\usepackage{mathrsfs}%
\usepackage[title]{appendix}%
\usepackage{xcolor}%
\usepackage{textcomp}%
\usepackage{manyfoot}%
\usepackage{booktabs}%
\usepackage{algorithm}%
\usepackage{algorithmicx}%
\usepackage{algpseudocode}%
\usepackage{listings}%
\usepackage{xcolor}
\usepackage{float}
\usepackage{lineno}

\makeatletter
\def\Authorfont{\reset@font\fontsize{8bp}{10bp}\selectfont\boldmath\titraggedcenter}%
\def\addressfont{\reset@font\fontsize{7bp}{9bp}\selectfont\titraggedcenter}%
\makeatother

\setcitestyle{super,sort&compress,open=,close=}
\begin{document}

\title{A Photonic Integrated Light Generation, Control, and Delivery Engine for Creating Cold Atoms}

\author[1]{\fnm{Andrei} \sur{Isichenko}\textsuperscript{\dag}}
\author[1]{\fnm{Calisto} \sur{Ruiz}}
\author[1]{\fnm{Chuan} \sur{Zhong}}
\author[1]{\fnm{Nick} \sur{Montifiore}}
\author[2]{\fnm{Sam J.} \sur{Harding}}
\author[1]{\fnm{Meiting} \sur{Song}}
\author[1]{\fnm{Chenghang} \sur{Liu}}
\author[1]{\fnm{Nitin} \sur{Indukuri}}
\author[3]{\fnm{Karl D.} \sur{Nelson}}
\author[1*]{\fnm{Daniel J.} \sur{Blumenthal}}

\affil[1]{\orgdiv{Department of Electrical and Computer Engineering}, \orgname{University of California, Santa Barbara}, \orgaddress{\city{Santa Barbara}, \postcode{93106}, \state{CA}, \country{USA}}}
\affil[2]{\orgdiv{Quantum Engineering Technology Labs, H.H. Wills Physics Laboratory and School of Electrical, Electronic, \& Mechanical Engineering}, \orgname{University of Bristol}, \orgaddress{\street{Tyndall Avenue}, \city{Bristol}, \postcode{BS8 1FD}, \country{UK}}}
\affil[3]{\orgname{Honeywell Aerospace Technologies}, \orgaddress{\city{Plymouth}, \state{MN}, \country{USA}}}
\affil[*]{Corresponding author: Daniel J. Blumenthal (\href{mailto:danb@ucsb.edu}{danb@ucsb.edu})}

\presentaddress{\textsuperscript{\dag} Infleqtion, Louisville, CO 80027, USA.}

\abstract{The creation of cold atoms underpins quantum sensing and precision metrology \cite{kitching_atomic_2011,meyer_assessment_2020,aveline_observation_2020,degen_quantum_2017,ye_quantum_2008}, computing \cite{graham_multi-qubit_2022,bluvstein_quantum_2022,evered_high-fidelity_2023}, timekeeping \cite{ludlow_optical_2015,norcia_seconds-scale_2019,pedrozo-penafiel_entanglement_2020}, and fundamental science experiments \cite{cairncross_atoms_2019,kaufman_quantum_2021}. Miniaturizing the lasers and optical infrastructure, which occupy over 80\% of the volume of these experiments, through photonic integration could dramatically improve robustness, reliability, and manufacturability while reducing cost and enabling portable space-based precision measurements \cite{burt_demonstration_2021,zhang_compact_2018,schuldt_design_2015} and  quantum tools for non-experts. Yet, there has been limited progress towards integration of the laser and optical infrastructure required to generate cold atoms onto a single chip. We report a significant advance in miniaturization of cold atom experiments with single-chip integration of a photonic engine that generates, controls, and delivers precision light to create cold atoms in a rubidium 3-dimensional magneto-optical trap (3D-MOT). We integrate on a single chip two frequency-agile self-injection-locked lasers, laser frequency and intensity control, passive beam combining and splitting, signal monitoring, and high-efficiency beam delivery to a rubidium vapor cell. The integrated engine is shown to support multiple functions required for generating cold atoms including stable generation of low-noise cooling and repump light, frequency control for atom spectroscopy and  stabilization, real-time laser monitoring, routing for laser offset locking, optical power modulation, and large-area three-beam delivery of 3 mW to the physics package. Fully electrically controlled, the engine enables sustained 3D-MOT operation, trapping more than 1 million atoms at an average temperature of 350~$\mu$K and supporting atom-number and temperature measurements. These results demonstrate the first single-chip integration of light generation, control, and delivery for cold-atom generation, offering a path toward smaller, more robust, and manufacturable cold-atom systems for applications including cold-atom clocks and sensors.}

\maketitle

\section*{Introduction}

The precise generation and control of laser-cooled neutral atoms underpins some of today's most sensitive measurement technologies and quantum systems, including optical atomic clocks \cite{ludlow_optical_2015,norcia_seconds-scale_2019,pedrozo-penafiel_entanglement_2020,burt_demonstration_2021}, atom-interferometric inertial sensors \cite{kitching_atomic_2011,aveline_observation_2020,schuldt_design_2015,zhang_compact_2018,degen_quantum_2017,kaufman_quantum_2021}, and platforms for quantum simulation and computation \cite{bluvstein_quantum_2022,evered_high-fidelity_2023,graham_multi-qubit_2022}. Realizing an ensemble of cold atoms requires the coordinated operation of multiple subsystems, including ultra-low frequency noise stabilized laser sources, precise frequency and intensity control, spatially uniform beam delivery to a vacuum containing atoms, and the proper alignment of optical beams with a quadrupole magnetic field \cite{blumenthal2024Enabling,monroe_very_1990}. The end-to-end functions from light generation to light delivery to the atomic ensemble are often referred to as a beamline. In conventional laboratory implementations, these beamlines are constructed using table- and rack-scale technologies and discrete optics and components connected using optical fibres and free-space components, ultimately delivering light to a vacuum chamber containing the atomic ensemble, referred to as the physics package. These lab-scale configurations limit the deployment of cold-atom technology outside the laboratory, its use by non-experts, its scaling and robustness, and the real-world use in a broader class of applications.

These constraints have motivated a sustained effort to miniaturize every stage of the cold-atom generation beamline and physics package \cite{blumenthal2024Enabling, mcgilligan2022Microfabricated}. Grating magneto-optical traps (GMOTs) \cite{nshii2013surfacepatterned, lee2022compact} and micro-fabricated vacuum cells \cite{hummon2018Photonic} have reduced the physics package to chip- and wafer-scale volumes, while chip-scale and hybrid-integrated semiconductor lasers \cite{kiewiet2026Heterogeneously, corato-zanarella2026LowSWaPa, isichenko_sub-hz_2024} have begun to replace bench-top light sources. This miniaturization has been enabled by ultra-high-$Q$ integrated resonator-based lasers such as self-injection-locked (SIL) semiconductor lasers, integrated extended cavity tunable lasers (ECTLs), and stimulated Brillouin scattering lasers~\cite{chauhan2026Chip} that provide self-isolation and co-integration with other photonic components such as integrated stabilization cavities and optical filters \cite{heim_hybrid_2025,heim_narrow-linewidth_2026, heim_versatile_2026}. These integrated laser technologies are capable of delivering linewidth, frequency noise, and stability comparable to and sometimes exceeding that of tabletop external-cavity diode lasers (ECDLs) traditionally used in cold atom experiments. Importantly, these integrated laser sources are agile and can be directly frequency tuned, enabling cold atom functions \cite{isichenko_sub-doppler_2026} without bulky, power consuming acousto-optic modulators. Critically, co-locating frequency agile, low frequency noise light generation on the same chip with other beamline components including passive optics, modulators and beam-delivery optics on a common substrate confers further advantages including lowering the laser output power requirements, suppression of accumulated phase noise inherent in lab-scale systems, reduced sensitivity to environmental vibration and disturbances, and reduced volume, weight and power consumption. The improvements in beam-pointing stability and phase noise accumulation can lead to enhanced MOT stability in the presence of environmental disturbances and sources of decoherence \cite{niffenegger2020Integrated}. Low-loss photonic integrated circuits, most notably the CMOS compatible silicon nitride platform \cite{blumenthal2018Silicon}, can deliver the light needed for laser cooling directly to trapped ions and neutral atoms using off-chip laser sources \cite{niffenegger2020Integrated,isichenko2023Photonica,mcgehee2021Magnetooptical,ferdinand2025Lasercooling}. What has remained elusive is demonstration of a platform that unifies light generation, control, and delivery into one compact system.

Here we demonstrate a packaged photonic integrated cold-atom light engine (PICMOT) that combines laser generation, stabilization, optical control and free-space beam delivery on a single silicon nitride chip to produce laser-cooled neutral atoms. The platform integrates two hybrid-integrated, frequency-agile self-injection-locked semiconductor lasers for cooling and repumping, together with on-chip frequency and intensity control, optical routing and splitting, and three free-space beam emitters that form a six-beam three-dimensional magneto-optical trap (3D-MOT) through retroreflection. The integrated lasers provide the frequency and intensity stability required for sustained MOT operation and repeated experimental cycles: locked to a Rb reference, the cooling-laser frequency fluctuates by 12.9 kHz r.m.s. over one hour, and the on-chip intensity servo stabilizes the 3 mW total trapping-beam power to 0.24 $\mu$W r.m.s. Using this architecture, we trap up to $1.12\times10^{6}$ $^{87}$Rb atoms at a temperature of $350~\mu\mathrm{K}$, demonstrating the coordinated operation of the integrated optical functions required for cold-atom generation. The packaged PIC occupies only $6.3\times5.3\times0.15~\mathrm{cm}^{3}$ and operates within a compact $20\times20\times20~\mathrm{cm}^{3}$ cold atom experimental assembly. By replacing much of the conventional free-space optical beamline with a lithographically defined, wafer-scale manufacturable photonic circuit, this work establishes a path towards compact, robust and scalable cold-atom systems that can be deployed beyond specialized laboratories and operated by non-experts.

\section*{Results}

\begin{figure}[h]
\centering
\includegraphics[width=1.0\textwidth]{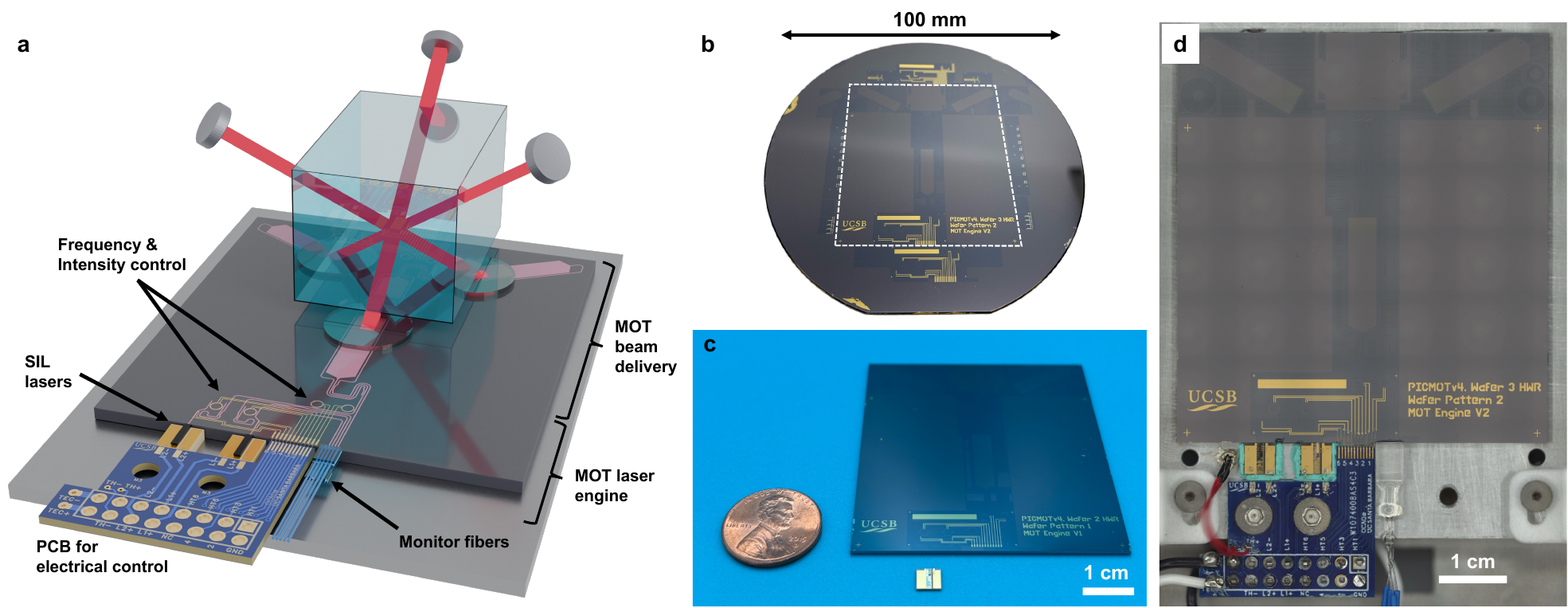}
\caption{\textbf{Creating cold atoms with a photonic integrated light generation, control, and delivery engine.} \textbf{a}, Rendering of the packaged light engine assembly showing frequency-agile cooling and repump hybrid-integrated SIL lasers, passive splitters, taps, and monitor ports, fiber v-groove array for saturated absorption spectroscopy, beat-note locking, and signal monitoring, frequency control traces, intensity control bus-ring modulator and control traces, and MOT 3-beam large-area delivery emitters. Also shown are the quarter-wave plates for circular polarization, rubidium vacuum vapor cell where the MOT is formed, and the retroreflecting mirrors that create the 6-beam configuration. The magnetic field coils are omitted for clarity. 
\textbf{b}, Photograph of a full silicon nitride wafer before dicing; the dashed box outlines one MOT engine PIC.
\textbf{c}, Photograph of pre-packaged PICMOT PIC and one of the SIL laser DBR chip-on-submount devices with penny for scale.
\textbf{d}, Photograph of packaged PICMOT light engine with wire-bonded lasers, electrical control PCB, and fiber v-groove array block for saturated absorption spectroscopy, beat-note locking, and signal monitoring (silicon nitride waveguides are enhanced for clarity). MOT: magneto-optical trap. PIC: Photonic integrated circuit. SIL: Self-injection locked. PCB: Printed circuit board.}
\label{fig:PICMOT_CHIP}
\end{figure}

The photonic integrated cold-atom light engine is a packaged hybrid-integrated silicon nitride (Si$_3$N$_4$, hereafter SiN) photonic integrated circuit (PIC) chip as shown in Fig.~\ref{fig:PICMOT_CHIP}. A 3D rendering (Fig. \ref{fig:PICMOT_CHIP}a) identifies the main photonic circuits including the self-injection locked (SIL) cooling and repump lasers, light frequency and intensity control, passive optical splitting and routing, routing to monitoring fibers, and large-area three-beam delivery to the 3D-MOT. Also shown are the electrical circuit board connections for control and the packaged fiber v-groove array for routing to off-chip monitoring of lasers, rubidium spectroscopy referencing, beat-note locking, and photodetection. The rendering also highlights the other main components of the 3D-MOT experiment, including quarter-wave plates for polarization conversion, three-mirror retroreflectors, and the rubidium vacuum cell. The quadrupole magnetic field coils are discussed in the Methods. Figure~\ref{fig:PICMOT_CHIP}b shows a photo of the fully processed 100-mm wafer with the PICMOT prior to dicing and Fig.~\ref{fig:PICMOT_CHIP}c shows the diced PICMOT chip and one of the distributed Bragg reflector (DBR) laser chip-on-submount devices with a penny for scale. The fully packaged hybrid-integrated PICMOT is shown in Fig.~\ref{fig:PICMOT_CHIP}d with only electrical inputs and fiber and free-space optical outputs. 

\begin{figure}[h]
\centering
\includegraphics[width=1.0\textwidth]{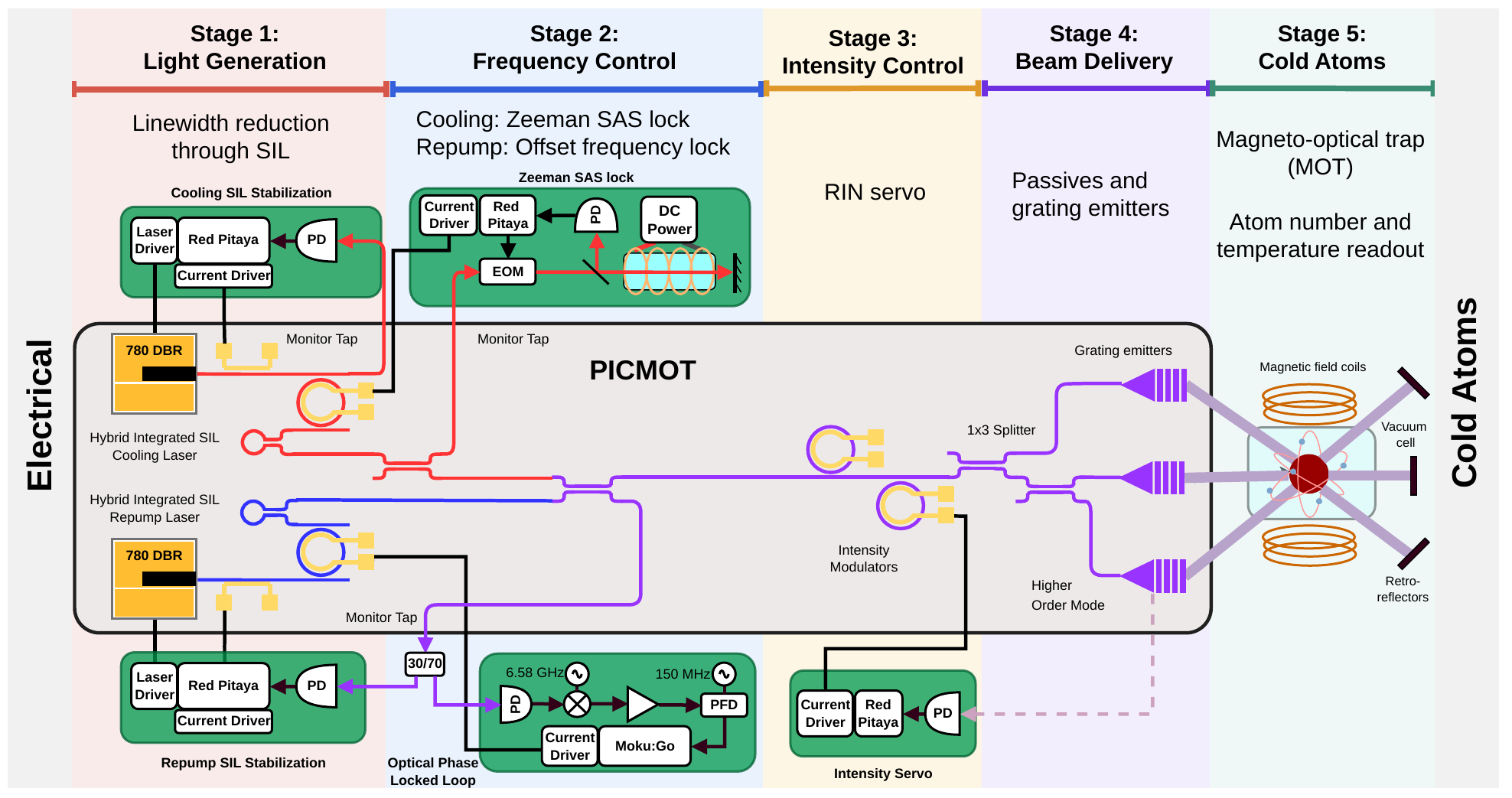}
\caption{\textbf{Experimental sequence architecture and integrated cold atom light engine PIC and experiment.}
Functional architecture of the PICMOT system, organized into five stages from light generation to cold-atom trapping. Stage 1, light generation and linewidth reduction: hybrid-integrated 780-nm DBR lasers are self-injection locked (SIL) to on-chip ring resonators to generate the cooling and repump light. Stage 2, frequency control: the cooling laser is stabilized to the $^{87}$Rb D$_2$ transition using Zeeman-tunable saturated-absorption spectroscopy (SAS), while the repump laser is offset-locked relative to the cooling laser at 6.58~GHz using an optical phase-locked loop (OPLL). Integrated monitor ports provide signals for off-chip feedback circuits. Stage 3, intensity control: integrated ring modulators and a relative-intensity-noise (RIN) servo stabilize the optical power delivered to the atoms via feedback from a free-space beam monitor signal. Stage 4, beam delivery: the combined cooling and repump light is split and routed to three large-area grating emitters that generate the three free-space trapping beams. Stage 5, cold atoms: the three emitted beams enter the $^{87}$Rb vacuum cell and are retroreflected to form the six-beam geometry for magneto-optical cooling and trapping in the quadrupole magnetic field. Electrical control and feedback coordinate the five stages and experimental sequence.
}
\label{fig:architecture}
\end{figure}

\subsection*{Cold atom light engine}

The light engine and the cold-atom sequence are organized into five stages as shown in Fig.~\ref{fig:architecture}. In Stage 1, a pair of hybrid-integrated self-injection-locked (SIL) lasers generates narrow linewidth and low frequency  noise cooling and repump light. In Stage 2, the cooling laser is scanned across the $^{87}$Rb saturated-absorption spectroscopy (SAS) spectrum to identify the atomic transitions and is then locked to the $^{87}$Rb $D_2$ cooling transition. The repump laser is offset-locked to the cooling laser via their beat-note. In Stage 3, an integrated thermo-optic intensity modulator~\cite{zhang2022240} controls and stabilizes the cooling-light intensity delivered to the atoms~\cite{nelson_relative_2008}. In Stage 4, the cooling and repump light is split equally among three waveguides and routed to grating emitters that launch the trapping beams into the physics package. In Stage 5, the six-coil quadrupole magnetic field~\cite{comptonMagneticFieldCoils2014} is sequenced with Stages 1--4 to load the MOT, and the atoms are characterized by fluorescence (atom number) and time-of-flight (TOF) absorption imaging (temperature and atom number).

Figure~\ref{fig:architecture} shows the PIC and its control electronics in more detail, with the PIC in the central grey panel and the control subsystems (green) arranged by stage. Two hybrid-integrated 780\,nm semiconductor distributed Bragg reflector (DBR) lasers are each self-injection locked to a SiN ring-resonator feedback circuit. The frequency of each SIL laser is controlled through the DBR injection current and through thermo-optic heaters (gold) on the phase shifter and the add-drop ring resonator. Identical cooling and repump SIL stabilization subsystems (Stage~1) hold each laser at its SIL operating point by feedback to the DBR current (see Stage~1 and Methods). The cooling laser is referenced to the $^{87}$Rb D$_2$ transition by a Zeeman-tunable saturated-absorption spectroscopy (Zeeman-SAS) lock~\cite{dyer2023ChipScale} (Stage~2, top). The cooling laser signal is extracted from the PICMOT engine via an on-chip monitor tap (output power 0.7 mW) and then sideband modulated with an electro-optic modulator (EOM) for locking to a $^{87}$Rb cell. The error signal is fed back to the ring heater of the cooling SIL laser to correct long-term frequency drift and a coil around the vapor cell Zeeman-shifts the SAS reference to set the red detuning of the cooling light. The cooling and repump outputs are then combined on chip, with a monitor tap of the combined light feeding an optical phase-locked loop (OPLL; Stage~2, bottom) that offset-locks the repump laser to the cooling laser by actuating the repump ring heater. The beat frequency is set to 6.58 GHz to match the $F=1 \rightarrow F'=2$ transition. The combined light then passes through two thermo-optic ring-resonator intensity modulators (Stage~3). An intensity servo stabilizes the trapping-beam power by measuring the otherwise unused $m=2$ beam of one grating emitter and actuating one of the ring heater modulators, which also suppresses relative intensity noise (RIN) (see Stage~3 and Methods). The cooling and repump light is then split three ways and delivered to the grating emitters (Stage~4). The SIL, SAS, and intensity locks run on Red Pitaya FPGA controllers~\cite{redpitaya2026}, and the OPLL runs on a Moku:Go~\cite{mokugo2026} (Methods). In future implementations, the SAS reference could be integrated on the PIC using similar technologies~\cite{hummon2018Photonic, dyer2023ChipScale}, and the atoms in the physics package could themselves serve as the frequency reference, with an integrated Si$_3$N$_4$ coil resonator providing holdover~\cite{chauhan2026Chip, isichenko_sub-doppler_2026}.

\begin{figure}[h]
\centering
\includegraphics[width=1.0\textwidth]{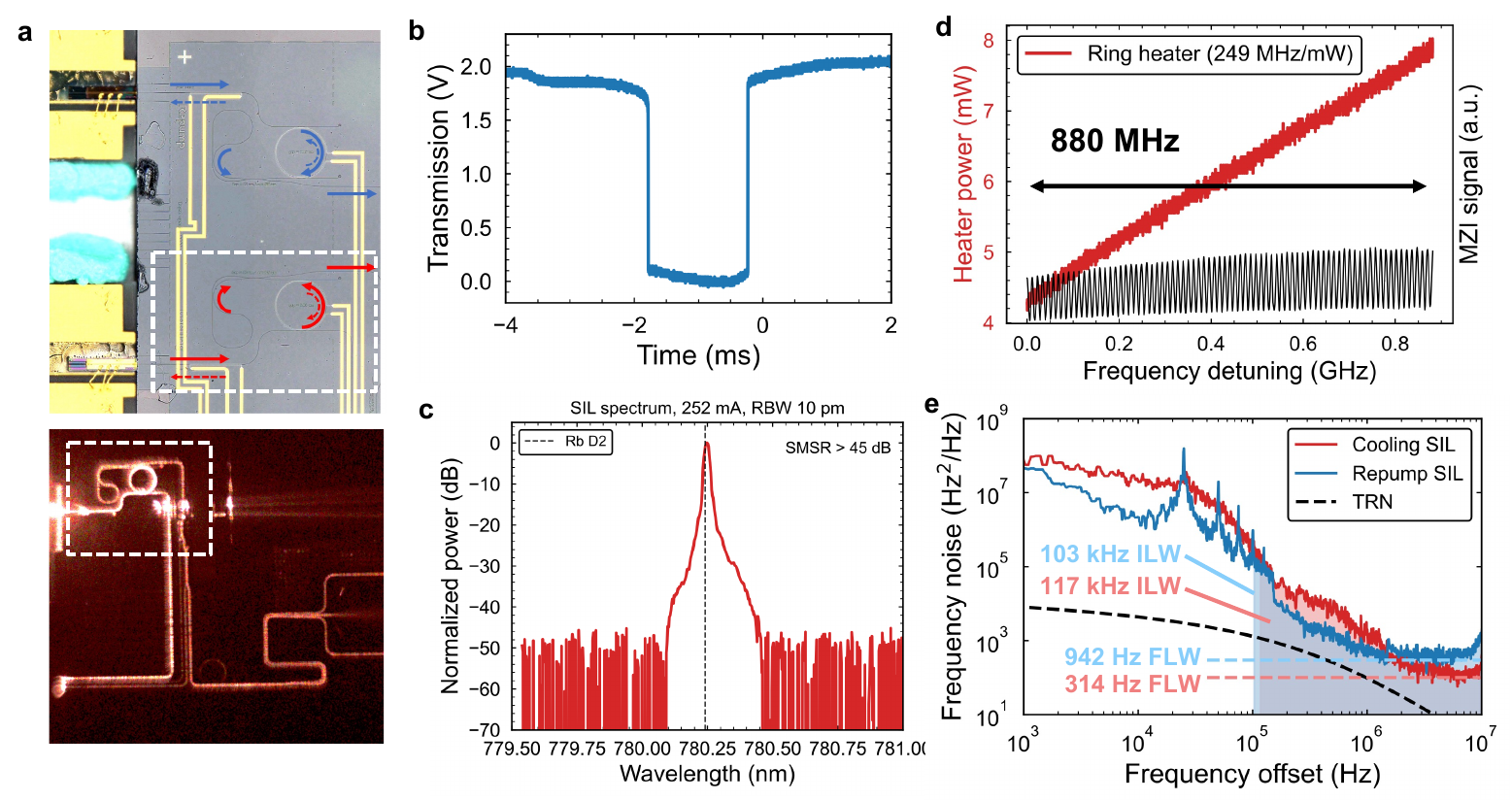}
\caption{\textbf{Integrated self-injection-locked (SIL) laser subsystem for cooling and repump light generation, frequency-noise reduction and agile tuning.}
\textbf{a,} Top: photograph of the packaged cooling and repump SIL lasers, with the DBR lasers (left) edge-coupled to the PIC. Arrows trace the light paths of the repump (blue) and cooling (red) lasers through the ring resonators and Sagnac loop mirrors (dashed: back-reflected light). Bottom: top-down scattered-light image of a feedback circuit illuminated with fiber-coupled 780 nm light. Dashed boxes mark the ring resonator and loop mirror in each image.
\textbf{b,} Ring-resonator transmission versus time during a sweep of the DBR current. The sharp edges mark the self-injection-locking range; the left edge provides the error signal for the side-of-fringe servo that holds the laser in the SIL state (Methods).
\textbf{c,} Optical spectrum of the cooling SIL laser at a 252 mA drive current (resolution bandwidth 10 pm) near the $^{87}$Rb $D_2$ transition (dashed line), showing single-mode operation with a side-mode suppression ratio greater than 45 dB.
\textbf{d,} Frequency tuning with the integrated ring heater. Heater power (red) and Mach--Zehnder interferometer (MZI) fringes (black, frequency calibration) are shown as functions of detuning, demonstrating continuous tuning over 0.9\,GHz with a coefficient of 249\,MHz\,mW$^{-1}$.
\textbf{e,} Frequency noise of the free-running cooling (red) and repump (blue) SIL lasers, together with the calculated thermorefractive-noise (TRN) floor of the resonator (dashed). The cooling laser reaches an integral linewidth (1/$\pi$ integration) of 117 kHz and a fundamental linewidth of 314 Hz; the repump laser reaches 103 kHz and 942 Hz, respectively.}
\label{fig:SIL_lasers}
\end{figure}

\clearpage
\subsection*{Stage 1: SIL Laser Light Generation}

The cooling and repump lasers are hybrid-packaged self-injection-locked (SIL) lasers. Each consists of a 780\,nm distributed Bragg reflector (DBR) laser edge-coupled to a low-loss Si$_3$N$_4$ feedback circuit (Methods). Figure~\ref{fig:SIL_lasers}a shows a photograph of the packaged chip and a top-down scattered-light image of the feedback circuit illuminated with fiber-coupled light, taken before integration of the DBR lasers. In the circuit, light from the DBR passes through a thermo-optic phase shifter, which sets the feedback phase, to a thermally tunable add-drop ring resonator of 750~$\mu$m radius, with an intrinsic quality factor of 2.5 million and a loaded quality factor of 0.4 million (strongly over-coupled). A Sagnac loop mirror with 10\% reflection at the drop port returns the ring-filtered light to the DBR. The loop mirror sets the feedback strength and the ring provides frequency selectivity, so the lasing frequency is set by the ring resonance and follows it as the ring is tuned.

Sweeping the DBR current while monitoring the ring transmission through an on-chip tap reveals the characteristic transition into and out of the SIL state~\cite{kondratiev2017Selfinjection} (Fig.~\ref{fig:SIL_lasers}b). The width of this region in the current sweep reflects the range of free-running laser detuning over which SIL is maintained. To hold each laser in the SIL state over long periods, a side-of-fringe servo acts on the DBR current: a fixed offset is subtracted from the tapped ring transmission so that the error signal crosses zero on the low-current edge of the SIL dip, and the servo holds the laser at this point so that it tracks the ring resonance despite slow thermal drift (Methods). Because the servo requires no modulation, it adds no sidebands to the light delivered to the atoms.

\begin{figure}[H]
\centering
\includegraphics[width=1.0\textwidth]{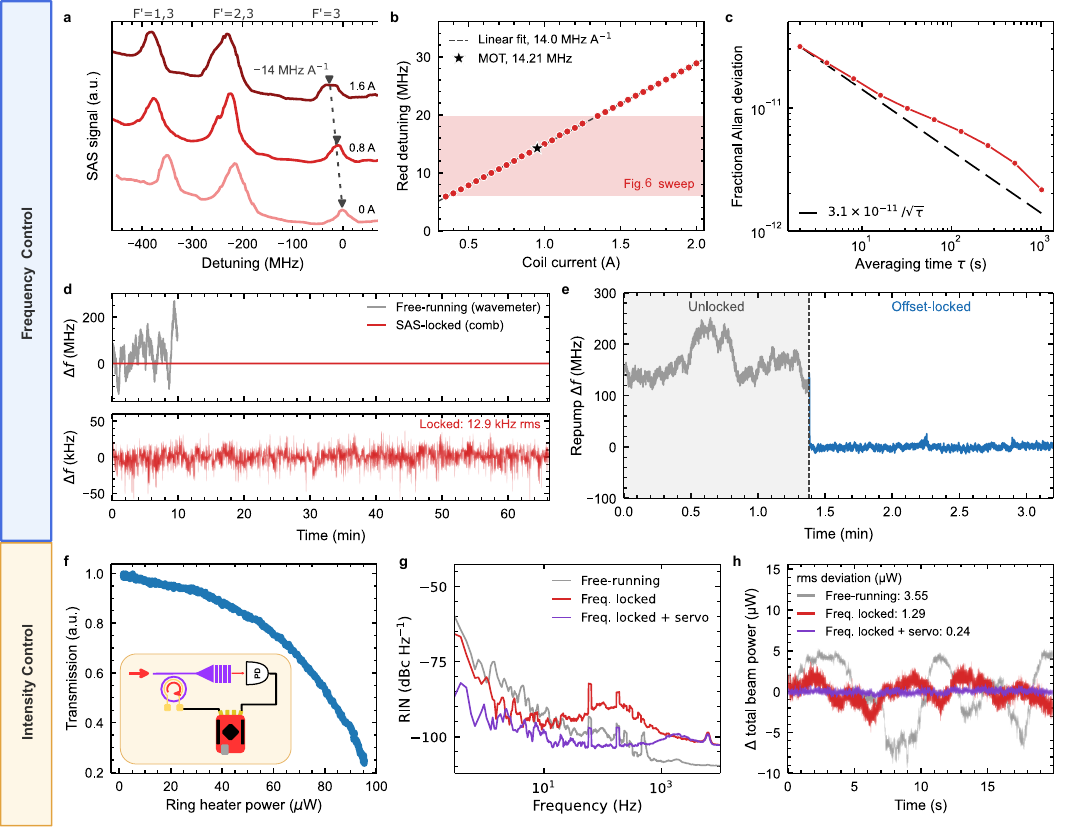}
\caption{\textbf{Laser frequency and power control.}
\textbf{a,} Saturated-absorption spectroscopy of the $^{87}$Rb $D_2$ $F=2$ manifold at Zeeman coil currents of 0, 0.8 and 1.6\,A (vertically offset), referenced to the zero-field $F=2\rightarrow F'=3$ transition. Triangles mark the lock points, red-shifted by 14.0\,MHz\,A$^{-1}$.
\textbf{b,} Cooling-laser red detuning versus Zeeman coil current (red), with a linear fit (14.0\,MHz\,A$^{-1}$, dashed).
\textbf{c,} Fractional overlapping Allan deviation of the locked cooling laser, calculated from the comb beat-note in \textbf{d}. Dashed line: $3.1\times10^{-11}\tau^{-1/2}$ scaling.
\textbf{d,} Cooling-laser frequency. Top: free-running (grey, wavemeter) and SAS-locked (red, ULE-referenced frequency comb). Bottom: expanded locked trace showing 12.9 kHz r.m.s. fluctuations over 65 min.
\textbf{e,} Repump-laser frequency measured by wavemeter before (grey) and after (blue) engaging the offset-frequency lock to the cooling laser (dashed line).
\textbf{f,} Normalized transmission through the on-chip intensity modulator versus ring-heater electrical power, with experimental schematic.
\textbf{g,} Relative intensity noise (RIN) spectra for three operating conditions.
\textbf{h},~Change in total MOT beam power for the free-running laser (grey), with the frequency lock engaged (red), and with both the frequency lock and the on-chip intensity servo engaged (purple). The r.m.s. fluctuations are 3.55, 1.29 and 0.24\,$\mu$W (0.12\%, 0.043\% and 0.008\%), respectively. 
}
\label{fig:control}
\end{figure}

The cooling laser operates at the Rb D$_2$ wavelength 780.24 nm with a side-mode suppression ratio (SMSR) above 45\,dB (Fig.~\ref{fig:SIL_lasers}c). With the DBR following the ring, the laser frequency can be steered with the ring heater alone, which is the actuator used for atomic referencing (Fig.~\ref{fig:control}a). Figure~\ref{fig:SIL_lasers}d shows continuous triggered scanning over 0.9\,GHz with a tuning coefficient of 249\,MHz\,mW$^{-1}$ of heater power. The frequency noise of both lasers is shown in Fig.~\ref{fig:SIL_lasers}e. The cooling laser reaches an integral linewidth (1/$\pi$ integration) of 117\,kHz and a fundamental linewidth of 314\,Hz; the repump laser reaches 103\,kHz and 942\,Hz, respectively. Both integral linewidths are more than 50 times narrower than the natural linewidth of the cooling transition, $\Gamma/2\pi = 6.07$\,MHz.

\subsection*{Stages 2 and 3: Light Control}

Following narrow-linewidth light generation by the cooling and repump self-injection-locked (SIL) lasers, Stages 2 and 3 provide the frequency and intensity control required for laser cooling and trapping of $^{87}$Rb. These functions include atomic spectroscopy and frequency stabilization, offset locking between the cooling and repump lasers, and optical power control with relative intensity noise (RIN) suppression. We first sweep the laser frequency, probing the $^{87}$Rb $D_2$ hyperfine transitions as shown in the spectroscopy trace in Fig.~\ref{fig:control}a. The cooling laser is then locked to $F=2 \rightarrow F'=3$ hyperfine transition using feedback to the SIL ring tuner for frequency control as described in the Cold Atom Light Engine Section and Methods. To enable long-term, robust  MOT operation, the cooling laser and the offset-locked repump frequencies must be held constant at specific offsets from the respective atomic transitions. We utilize saturated-absorption-spectroscopy (SAS) for referencing the lasers to atomic transitions \cite{preston1996Dopplerfree}. An error signal is generated at each hyperfine transition, which is used to lock the laser frequency directly to the transition as described in further detail in Methods. In order to form a MOT, the cooling laser is red-detuned from the $F=2 \rightarrow F'=3$ cooling transition by $\sim$10-20 MHz, outside of the nominal SAS error signal capture range. To bring the SAS signal to within range and enable MOT formation we employ a Zeeman-tunable SAS lock. By applying a magnetic field to the $^{87}$Rb vapor cell used for SAS (configuration shown at the top of Stage 2 in Fig.~\ref{fig:architecture}), the cooling transition energy is shifted through the Zeeman effect \cite{ho_seong_lee_frequency_1996}. The transition and the corresponding lock point are dynamically tuned with current through the coils to achieve the desired detuning from the cooling transition (see tuning curve in  Fig.~\ref{fig:control}b). The SIL cooling laser frequency is controlled directly by feeding back to resistive heaters on the SIL ring, allowing for rapid frequency corrections. In the future this function could be integrated using a miniature MEMS vapor cell with PCB coils~\cite{dyer2023ChipScale}, compatible with our large-area SiN beam delivery.

The frequency of the Zeeman-tuned, SAS-locked cooling laser is measured with a heterodyne beat-note with a self-referenced fiber frequency comb, frequency-doubled to 780\,nm and stabilized to a 1\,Hz-linewidth ultra-low-expansion (ULE) cavity (Methods). The beat-note shows 12.9 kHz r.m.s. frequency fluctuations over 65 min (Fig.~\ref{fig:control}d), with an overlapping Allan deviation of $3.1\times10^{-11}/\sqrt{\tau}$ (Fig.~\ref{fig:control}c). Because the MOT fully loads over 0.5 seconds (Fig.~\ref{fig:MOT}d), laser frequency drift is negligible over a loading cycle, and the residual fluctuations remain $\approx2\times10^{-3}$ of the natural linewidth $\Gamma$. In steady-state operation, with all locks engaged, the MOT runs continuously without re-acquiring any lock.

The absolute frequency and stability of the atomically referenced cooling laser are transferred to the repump laser through a 6.58 GHz optical phase-locked loop (OPLL), shown at the bottom of Stage 2 in Fig.~\ref{fig:architecture}. The repump laser frequency offset from the cooling laser is set by a low-noise reference electronic oscillator, allowing for the repump frequency to be set exactly at the repump transition without an additional SAS lock. Additionally, the repump laser frequency can be dynamically tuned across hundreds of megahertz by changing the reference VCO frequency. The wide and continuous tunability is useful for finding optimal conditions for MOT formation. Engaging the offset lock suppresses the repump frequency excursions from over 100\,MHz to near the wavemeter resolution (Fig.~\ref{fig:control}e) as the repump tracks the cooling laser stability.

In addition to laser frequency control, we demonstrate programmable control of the trapping-beam intensity and RIN reduction using a thermally actuated ring intensity modulator. The modulator sets the power delivered to the atoms through the digital control interface and, in closed loop with a photodetector monitoring a higher-order grating beam, acts as a RIN-suppression servo. The modulator provides a maximum extinction ratio of 6.3 dB (Fig.~\ref{fig:control}f), of which only 2.2 dB is used in operation because of the SIL lasers' sensitivity to back-reflection from the single bus-coupled ring modulator (see Discussion Section) while in the moderate to strong extinction state. Engaging the servo reduces the r.m.s.\ power fluctuation of the MOT beams from 1.29~$\mu$W with the frequency lock alone to 0.24~$\mu$W (0.008\% of the 3.0 mW total; Fig.~\ref{fig:control}h). In the frequency domain, the servo lowers the RIN by 10--15 dB between $\sim$1 Hz and 1 kHz, reaching $\approx-103$~dBc Hz$^{-1}$ (Fig.~\ref{fig:control}g). The thermal tuner frequency response sets the $\sim$2\,kHz loop bandwidth. Much of the suppressed noise in the 20 Hz -- 1 kHz band is added by the frequency lock itself, most likely as frequency corrections to the SIL ring heater are converted into amplitude fluctuations.

\begin{figure}[h]
\centering
\includegraphics[width=0.9\textwidth]{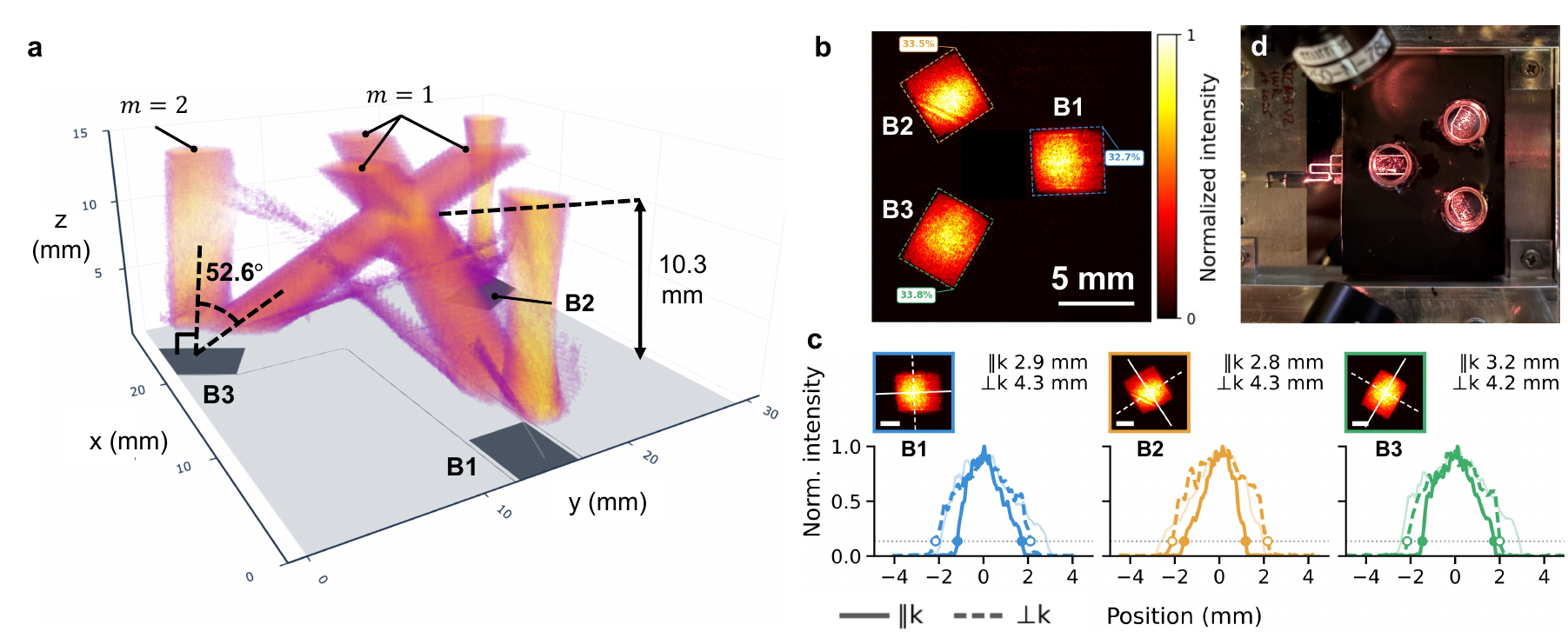}
\caption{\textbf{PICMOT beam delivery and characterization.}
\textbf{a,} Measured three-dimensional beam intensity distribution (B1--B3), reconstructed from beam profiles acquired at successive heights above the PIC. The trapping ($m=1$) and higher ($m=2$) diffraction orders, beam-intersection height, and trapping-beam angle are indicated.
\textbf{b,} Trapping-beam profiles above the intersection point. Fractional powers of B1, B2 and B3 are 32.7\%, 33.5\% and 33.8\%, respectively. Scale bar, 5\,mm.
\textbf{c,} Beam intensity profiles parallel ($\parallel k$) and perpendicular ($\perp k$) to the in-plane propagation direction. The measured $\parallel k$ profiles (light solid lines) are corrected by $\cos 52.6^\circ$ for the beam angle relative to the surface normal, yielding profiles normal to each beam (solid lines). Dashed lines show perpendicular profiles; markers indicate the $1/e^2$ beam diameters (dotted line). Scale bars, 2\,mm.
\textbf{d,} Fiber-coupled reference PIC installed in the MOT apparatus for independent beam-delivery characterization. Three quarter-wave plates above the emitters provide circular polarization and filter higher diffraction orders. The fiber-to-trapping-beam insertion loss is 9.4\,dB. Three quarter-wave plate--mirror retroreflectors generate the six-beam MOT geometry; B2 and B3 mounts are visible at left.}
\label{fig:Beam_delivery}
\end{figure}

\subsection*{Stage 4: Beam Delivery}

Following light generation, frequency stabilization, and intensity control, the cooling light is split and routed to three large-area integrated grating emitters (Fig.~\ref{fig:Beam_delivery}a). The splitter consists of a cascaded pair of $2\times2$ directional couplers designed for equal three-way splitting, and the routing waveguide width of 900~nm was chosen to reduce the risk of waveguide breakage and improve fabrication yield. From loss-spiral test structures on the same chip (Extended Data Fig.~\ref{fig:PIC_measurements}), we estimate a propagation loss of 1.7~dB over the $\sim$73-mm path from the splitter input to the emitters.

At each emitter, the routing waveguide expands into a slab region, from which the light is diffracted out of plane by a grating with a measured partial-etch depth of 18.5~nm (Methods, Extended Data Fig.~\ref{fig:AFMs}). Figure~\ref{fig:Beam_delivery}a shows the three-dimensional intensity distribution reconstructed from a vertical intensity scan. The grating centers lie on a circle of 13.5~mm radius, and each emits a collimated beam at 52.6$^{\circ}$ from the PIC surface normal, so that the three beams intersect 10.3~mm above the chip surface with an 87$^{\circ}$ angle between each pair. Each grating emits its trapping beam in the $m=1$ diffraction order, with a fraction of the light emitted in the $m=2$ order close to the surface normal. Figure~\ref{fig:Beam_delivery}b shows the PIC-emitted beams imaged above the beam intersection point. The three $m=1$ trapping beams, B1--B3, carry 32.7\%, 33.5\%, and 33.8\% of the total trapping-beam power, demonstrating near-equal power splitting among the emitters. The $m=1$ order accounts for 51\%, 57\%, and 57\% of the combined $m=1$ and $m=2$ power for B1--B3, respectively. The beams have average $1/e^2$ diameters of $w_{\parallel}=3.0$~mm in the emission plane and $w_{\perp}=4.3$~mm parallel to the chip surface (Fig.~\ref{fig:Beam_delivery}c), measured normal to the propagation direction (Methods), consistent with the previously reported design~\cite{isichenko2023Photonica}. The trapping beams show $\approx$6\% r.m.s. intensity variation over the central 20\% of the mode area and $\approx$15\% over the central 80\%.

The emitted $m=1$ beams are linearly polarized, so a quarter-wave plate (QWP) assembly positioned above the emitters converts them to circular polarization with the handedness required by the quadrupole magnetic field gradient. The same assembly also spatially filters out the $m=2$ orders. Three retroreflectors, each comprising a QWP and a mirror, then return the beams to form the six-beam three-dimensional MOT geometry (Methods). To characterize the beam-delivery network independently of the upstream stages, we fabricated a reference PIC containing only the beam-delivery circuit, into which externally generated light is fiber-coupled directly at the splitter input. Figure~\ref{fig:Beam_delivery}d shows this reference PIC installed in the MOT apparatus with the QWP assembly. On this chip, the total loss from the fiber input to the three trapping beams is 9.4~dB, compared with 16~dB for the previously reported beam-delivery PIC~\cite{isichenko2023Photonica}.

\subsection*{Stage 5: Cold Atom Generation and Measurements}

\begin{figure}[h]
\centering
\includegraphics[width=1.0\textwidth]{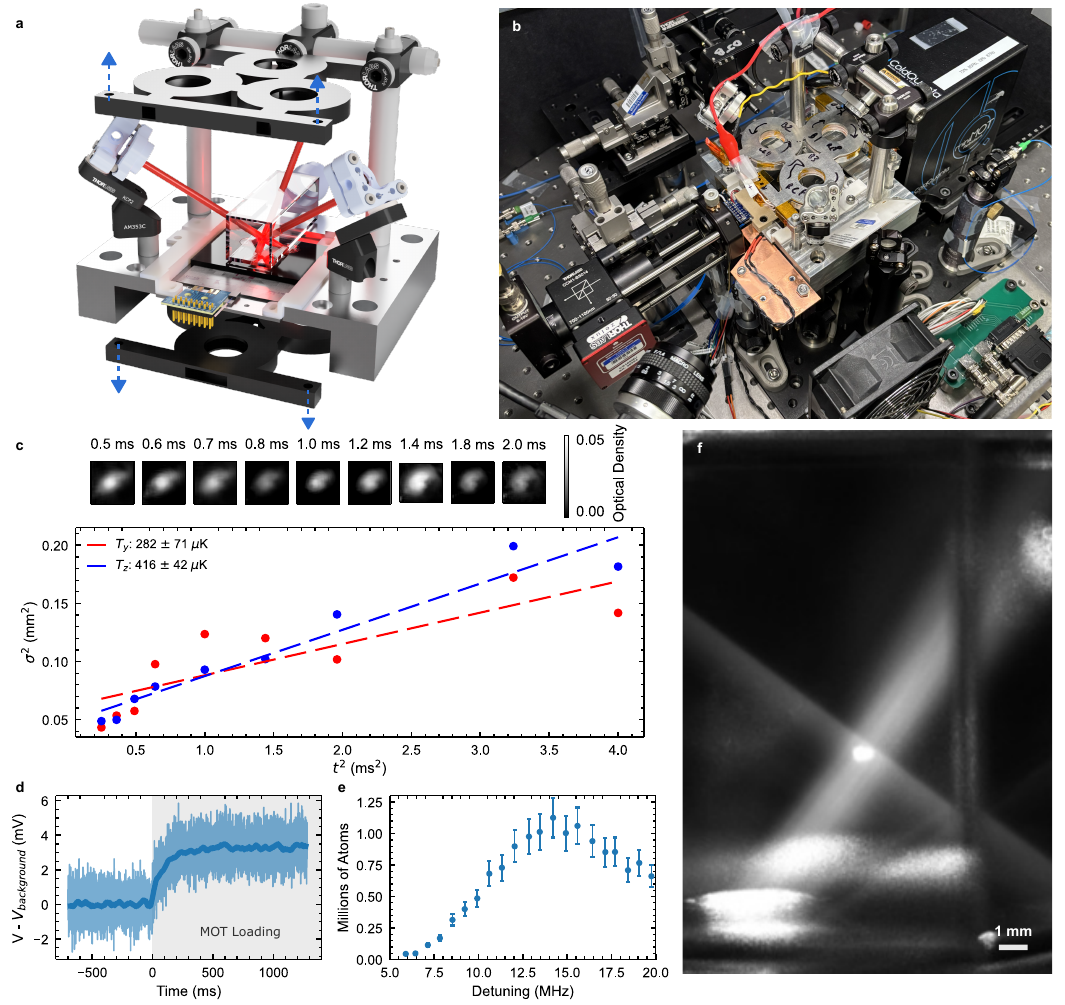}
\caption{\textbf{Rubidium MOT demonstration.}
\textbf{a,} Rendering of experimental setup with coils offset vertically for visibility.
\textbf{b,} Photograph of the PICMOT experimental setup, including the absorption and fluorescence imaging systems positioned above and to the left of the MOT physics package.
\textbf{c,} Cloud temperature characterization. Top: absorption images of the expanding cloud after release from the trap for increasing expansion times. Bottom: squared cloud widths $\sigma_y^2$ (red) and $\sigma_z^2$ (blue) plotted against squared expansion time $t^2$, with linear fits (dashed lines) to $\sigma^2(t) = \sigma_0^2 + (k_B T/m)\,t^2$. The fits give temperatures of $T_y = 282 \pm 71~\mu\mathrm{K}$ and $T_z = 416 \pm 42~\mu\mathrm{K}$. 
\textbf{d,} Photodiode signal during MOT loading, with the trap switched on at $t = 0$ (light trace: raw data; dark trace: averaged). At a detuning of $14.21$~MHz this corresponds to $N = (1.12 \pm 0.16)\times10^{6}$ atoms. 
\textbf{e,} Steady-state atom number versus cooling-laser detuning set by the Zeeman-tuned SAS lock. The atom number is maximized near $14.2$~MHz.
\textbf{f,} Side-view camera image of the glass vacuum cell, showing fluorescence from the trapping beams as they propagate through the rubidium vapor and intersect to form the MOT. The bright spot at the beam intersection corresponds to the trapped atom cloud.
}
\label{fig:MOT}
\end{figure}

Precisely aligned laser frequencies and beam balancing, combined with a magnetic field and a physics package, are the key ingredients to realizing a MOT, the first stage of creating cold matter. In our system, the magnetic field is generated by two sets of three in-plane coils, mounted on trays above and below the vacuum cell (Fig.~\ref{fig:MOT}a). The upper tray is rotated 180$^{\circ}$ relative to the lower tray to create a null in the magnetic field at the optical beam intersection point~\cite{comptonMagneticFieldCoils2014}. This configuration allows the optical beams to pass through to the retroreflectors while generating the magnetic field close to the trap site. We measure a magnetic field gradient of 21.5 G/cm along the beam B2 axis (Extended Data Fig.~\ref{fig:bfield_grad}, Methods).

The thermal atom source and vacuum cell used for the MOT are part of an Infleqtion miniMOT physics package (Fig.~\ref{fig:MOT}b). The package consists of a vacuum cell, Rb dispenser, and drive electronics. The vacuum cell is an ultra-high vacuum (UHV) glass cell with an internal ion pump sealed with the thermal atom source inside. The cell is not anti-reflection coated, with a measured transmission of 0.9 at 780~nm, adding another loss source for each pass through the cell. The atomic species makeup of the thermal source is 72\% $^{85}$Rb and 28\% $^{87}$Rb. The lower $^{87}$Rb content requires higher dispenser currents, which limits the achievable atom number. A different physics package with anti-reflection coatings and an $^{87}$Rb-enriched vapor cell would lower the required optical power.

We measure the MOT temperature using time-of-flight (TOF) absorption imaging~\cite{lett_observation_1988, ketterle_making_1999}. At the start of the TOF sequence, the magnetic quadrupole field and trapping light are rapidly switched off, releasing the atoms from the MOT. The cooling light is extinguished by setting the cooling DBR laser drive current to zero for 100~ms, while the repump laser remains on. After a free-expansion time $t_{\mathrm{TOF}}$, the atoms are illuminated from the side by a 300~$\mu$s probe pulse delivered through a free-space collimator (see Methods). The probe is generated by an off-chip laser frequency-stabilized using a separate rubidium spectroscopy module. Absorption images are acquired for a series of TOF values, with representative optical-depth (OD) images shown in Fig.~\ref{fig:MOT}c. Linear fits of the squared cloud widths versus $t_{\mathrm{TOF}}^2$ give temperatures of $T_y = 282 \pm 71~\mu$K and $T_z = 416 \pm 42~\mu$K (Fig.~\ref{fig:MOT}c).

\clearpage

We measure the MOT atom number using two independent methods: absorption imaging and fluorescence detection~\cite{hardmanBEC, rushton_novel_2015}. For absorption imaging, the atom number is extracted from the measured OD. As an independent measurement that does not require switching off the trapping light, we image the MOT fluorescence onto an amplified photodetector using a single collection lens. The atom number is determined from the detected fluorescence power using the collection solid angle and photon scattering rate, with the cooling-laser detuning independently measured from its beat-note with an external SAS-locked reference laser. Both techniques are described in Methods. Using fluorescence detection, we measure a maximum MOT population of $(1.12 \pm 0.16)\times10^6$ atoms at a cooling detuning of 14.2~MHz. The corresponding MOT loading curve is shown in Fig.~\ref{fig:MOT}d. Measurements of atom number as a function of cooling-laser detuning show a maximum near 14.2~MHz (Fig.~\ref{fig:MOT}e). Absorption imaging yields a peak atom number of approximately $0.5\times10^6$, about a factor of two below the fluorescence result. A side-view image of the MOT is shown in Fig.~\ref{fig:MOT}f, revealing fluorescence from the three trapping beams as they propagate through the rubidium vapor cell and intersect at the MOT location. The bright fluorescence at the beam intersection highlights the trapped atom cloud.

\section*{Discussion}

We have demonstrated a photonic-integrated light engine that generates, stabilizes, controls and delivers the optical fields required to cool and trap $^{87}$Rb. A single packaged silicon nitride PIC, driven by two hybrid-integrated semiconductor lasers, performs functions that conventionally require multiple lasers, frequency-shifting elements, modulators and free-space beam-delivery optics. To our knowledge, this represents the most complete integration of the optical infrastructure for cold-atom generation on a single packaged photonic circuit. Previous approaches have primarily miniaturized the downstream beam-delivery optics, using diffraction gratings~\cite{nshii2013surfacepatterned,lee2022compact}, metasurfaces and planar optics~\cite{mcgehee2021Magnetooptical}, or photonic integrated circuits driven by external laser systems~\cite{isichenko2023Photonica,ferdinand2025Lasercooling}. Here, low-loss waveguides extend this integration upstream to laser generation and control: high-$Q$ resonators self-injection lock and narrow the lasers, thermo-optic actuators control their frequencies and powers, and large-area grating emitters form the trapping beams. This combination allows the laser frequencies, optical powers and trapping geometry to be defined by the same photonic circuit and designed together with the atomic system.

The resulting engine provides cooling and repump light, linewidth narrowing, frequency tuning across the Rb hyperfine spectrum, atomic frequency stabilization, offset locking between the cooling and repump lasers, programmable intensity control and RIN suppression, optical monitoring, and large-area three-beam delivery to the vapor cell. Frequency-agile operations are performed by directly tuning the self-injection-locked lasers, eliminating acousto-optic modulators and their associated electrical power and free-space alignment. With all feedback loops engaged, the system maintains the cooling-laser frequency to 12.9 kHz r.m.s. over 65 min and stabilizes the trapping-beam power to within 0.24 $\mu$W r.m.s. The engine traps up to $1.1\times10^6$ atoms and integrates the frequency tuning and intensity control needed for polarization-gradient cooling (PGC), which requires coordinated laser power and detuning ramps together with magnetic-field control. Such cooling sequences have previously been demonstrated using rapidly tunable integrated resonators, reaching temperatures near $20~\mu$K~\cite{isichenko_sub-doppler_2026}. These results show that photonic integration can extend beyond passive beam delivery to encompass the frequency and intensity control required for a complete cold-atom experimental sequence.

The limitations of the present implementation arise from the actuators rather than the photonic architecture. The ring intensity modulator provides 6.3 dB of extinction, but only 2.2 dB can presently be used because tuning its resonance close to the laser frequency introduces resonant back-reflection that injection-locks the cooling laser and pulls it from the SAS lock point (Extended Data Fig.~\ref{fig:shutter_feedback}). We  attribute this behavior to the modulator having a higher loaded $Q$ than the cooling SIL resonator. A deliberately lower-$Q$ modulator or a Mach–Zehnder architecture~\cite{montifiore2026Blue} may suppress this interaction. The thermo-optic actuators also limit the intensity-control bandwidth to approximately 2 kHz, so full extinction during time-of-flight measurements currently requires switching off the cooling-laser current. Stress-optic actuators compatible with silicon nitride can provide MHz-scale bandwidth\cite{isichenko_sub-doppler_2026, mishra2026Ultralowa}, offering a route to fast, high-extinction on-chip switching without cycling the lasers.

Several functions remain outside the photonic package: polarization conversion and beam redirection, photodetection, the atomic frequency reference, magnetic-field generation and electronic control. None requires a change to the basic light-engine architecture. Photodetectors can be heterogeneously integrated with ultra-low-loss silicon nitride~\cite{tabatabaei_heterogeneously_2026,chawlani_heterogeneously_2026}, while waveplates and retroreflectors can potentially be replaced by planar metasurfaces~\cite{heki_high_2024,wang2026ThreeDimensional}. The external SAS reference could similarly be reduced to a chip-scale vapor-cell reference~\cite{dyer2023ChipScale,hummon2018Photonic}, or ultimately replaced by interrogation of the trapped atoms with a compact reference cavity providing holdover between measurement cycles~\cite{chauhan2026Chip}. Importantly, the light engine remains outside the vacuum system, allowing the photonics and atom source to evolve independently. The planar geometry also enables co-design with compact magnetic-field coils, which at present constitute the largest electrical power load of the experiment.

The measured atom number is likewise not a fundamental limit of the integrated architecture. MOT capture increases strongly with beam diameter~\cite{monroe_very_1990}, making larger emitters a direct route to increased capture volume. In the present layout, lowering the 10.3 mm beam-intersection height and feeding the emitters from inside the grating ring would shorten the waveguide routing and recover an estimated 1–2 dB of optical power. Suppressing higher diffraction orders, which presently carry $\sim$45\% of the emitted power, would provide a further increase in useful trapping power. Higher-power sources compatible with the same silicon nitride platform, including external-cavity tunable lasers~\cite{heim_hybrid_2025,nejadriahi2024Sub100} and extended-DBR lasers~\cite{siddharth2024Piezoelectrically}, provide an additional route to larger atom number. 

More broadly, integrating light generation, stabilization, control and beam delivery on a common platform changes how cold-atom systems can be constructed. The same architecture can incorporate frequency-agile beam pairs for atom interferometry~\cite{kodigala2024Highperformance}, optical tweezers and structured-light traps~\cite{graham_multi-qubit_2022,fang2026Metasurfaces}, or higher-power beams for evaporative cooling and Bose--Einstein condensation~\cite{aveline_observation_2020}. Rather than treating each cold-atom apparatus as a collection of independently aligned optical components, the PICMOT provides a lithographically defined light engine in which optical frequencies, powers and beam geometry are encoded in a manufacturable circuit. Such engines could be combined with different vacuum packages, replicated around a larger physics cell, or used to address multiple atom clouds. As neutral-atom quantum processors scale toward larger qubit arrays, increasing demands on optical power, thermal management and spatially resolved control may favor modular architectures that interconnect multiple atomic processors~\cite{saffman2026Neutrala,menssen2026Strategic}. These architectures place a premium on compact, reproducible optical subsystems that can be replicated across modules. The integrated light engine demonstrated here offers a path toward simplifying the atom-preparation infrastructure required for such systems, including independently operated cold-atom sources for replenishing computational arrays. This shift from assembled optical systems toward reproducible photonic subsystems offers a path to reducing the size, power, alignment complexity and unit-to-unit variation of cold-atom sensors, clocks and quantum processors.

\section*{Methods} 
\subsection*{Fabrication}

The fabrication process is designed to be compatible with CMOS-foundry processing (Extended Data Fig.~\ref{fig:Process_flow}). The starting substrate is a standard 4 inch diameter, 1 mm thick, Si wafer with a 15 $\mu$m thermal SiO$_2$ lower cladding. A 120 nm stoichiometric Si$_3$N$_4$ layer is deposited by low-pressure chemical vapor deposition (LPCVD) to serve as the waveguide layer. The Si$_3$N$_4$ is then patterned with a 248 nm deep-ultraviolet (DUV) ASML PAS 5500 stepper and etched using inductively coupled plasma (ICP) with optimized CHF$_3$/CF$_4$/O$_2$ chemistry. Following an O$_2$ plasma clean, a second lithography and etch sequence defines the shallow grating features. Lower RF power is used for the partial Si$_3$N$_4$ etch, with an average etch depth of 18.5~nm. After the final O$_2$ plasma clean, a 3 $\mu$m tetraethyl orthosilicate (TEOS) SiO$_2$ upper cladding is deposited by plasma-enhanced chemical vapor deposition (PECVD). The wafer is then annealed for 11 hours at 1100~$^\circ$C. Subsequent Ti/Pt (3~nm/250~nm) and Ti/Au (5~nm/1~$\mu$m) depositions are performed using electron-beam evaporation, with each deposition followed by metal lift-off to define the Pt heaters and Au pads for wire bonding. The heaters are designed with a 1 $\mu$m lateral edge-to-edge spacing with respect to the waveguide to reduce loss contribution. Two layers of photoresist are used during the metallization lithography steps to generate an undercut, enabling reliable metal lift-off across the 4 inch wafer and minimizing metal breakage. Finally, the wafer is diced to expose the chip facets for fiber and gain chip edge coupling. 

Critical waveguide dimensions are evaluated with scanning electron microscopy (SEM) after pattern transfer in step 3 of the process flow depicted in Extended Data Fig.~\ref{fig:Process_flow}. Measurements include the waveguide width at the input of the three beam expanders, routing-waveguide widths, directional coupler gaps, and inverse taper width at the dicing facet. Extended Data Table~\ref{SEMs_CD} summarizes the deviations in critical dimensions between the intended design and fabricated device, with SEM images shown in Extended Data Fig.~\ref{fig:SEMs}. The measurements are used to assess process variations and determine the minimum resolvable feature size from the photolithography, as well as to identify potential defects during the fabrication process. Critical dimensions for the partial-etch grating emitters are measured using atomic force microscopy (AFM) after pattern transfer in step 4 of the process flow, Extended Data Fig.~\ref{fig:Process_flow}. AFM images are acquired at three different locations on the grating and the corresponding height profiles are shown in Extended Data Fig.~\ref{fig:AFMs}. High uniformity and yield of sub-wavelength features across the wafer are enabled using an automated mask stitching protocol built with the python library, ASML\_JobCreator \cite{D_John_ASML_JobCreator_2024}, developed by Demis John at the UCSB Nanofab. The automated stitching also allows for the entire waveguide pattern, spanning the 4 inch wafer, to be condensed to two 21 mm by 25 mm photomasks.

Stringent particle control protocols are implemented throughout the fabrication process, with zero tolerance for particulate contamination along the beam delivery path. Surface cleanliness across the wafer is assessed using a Surfscan 6200 particle detector with a 488 nm laser. Using a 0.160 $\mu$m detection threshold and a 6 mm edge exclusion, 204 light-point defects (LPDs) were detected across the 4 inch wafer, corresponding to an average LPD density of 3.35~cm$^{-2}$. The mean detected defect diameter was $0.415 \pm 0.333~\mu$m, and approximately 78\% of the LPDs were smaller than 0.506~$\mu$m, indicating that the detected population was dominated by sub-micrometer features.

 Maintaining a clean backside of the wafer is crucial for uniformity across the beam-delivery section, which spans nearly the entire 4-inch wafer. Before each lithography step, the wafer backside is cleaned with acetone and isopropyl alcohol (IPA) while the wafer was spun face down on a POLOS spin coater. Remaining photoresist after solvent cleaning is  carefully removed mechanically with a razor blade. The backside is then visually inspected to confirm the absence of large particles and photoresist residue before loading into the ASML stepper. The procedure minimizes backside contamination that would otherwise alter the focus across the wafer during exposure.

Image quality control (IQC) protocol is implemented to yield the highest pattern quality when defining the photonic integrated circuits in this work. Exposure calibration is performed separately for the waveguide and partial-etch grating lithography steps. The calibration wafers consisted of Si$_3$N$_4$ on thermal SiO$_2$ on Si. The exposure doses were $14.5~\mathrm{mJ\cdot cm^{-2}}$ and $25~\mathrm{mJ\cdot cm^{-2}}$ for the two respective lithography steps with a focus offset of 0~$\mu$m. An IQC procedure is performed before each exposure to maintain an absolute focus correction below 20~nm.

\subsection*{Packaging}

\textit{Fiber-attach.} The monitor signals for all laser lock loops are routed out to their respective subsystems using a fiber V-groove array (VGA, OzOptics VGA-4-250-0-X-10.3-3.8-2.03-P-780-5/125-3A-1-1-0-Pyrex) attached to the PIC facet. The V-groove array is manipulated and aligned to the waveguide output taper using vacuum chucks mounted to a 3-axis stage (Thorlabs MAX312). Once optimally aligned, Norland Optical Adhesive 86 (NOA 86) is used to bond the V-groove array to the chip facet and a UV lamp (Dymax BlueWave 75) cures the adhesive. 

\textit{Laser-attach.} The SIL laser attach is performed using a custom-built, electronically-controlled, packaging system. An electrically actuated gripper (Schunk EGP 40-N-N-B) with custom fingers is used to pick up and manipulate the semiconductor lasers mounted on aluminum nitride (AlN) submounts. The lasers are driven with a PCB probe card using pogo pins. The laser is then coarsely aligned to the waveguide output taper using a 6-axis hexapod motion stage (PI H-811.S2) and fine adjustment is performed with piezo actuators on a 3-axis stage (Thorlabs MAX312). Once aligned, NOA 86 is deposited on the edges of the submount following the guidance of \cite{levaur2025Automated}. During all alignment, the laser is driven in pulsed mode with a 1 kHz repetition rate with 10 $\mu s$ pulse width (1\% duty cycle) to minimize heating while the laser is not temperature controlled. The adhesive is cured with a UV lamp, fixing the laser to the PIC facet. To heatsink the laser, thermal gap filler (Laird TPutty 607) is injected between the submount and the aluminum baseplate, shown in Extended Data Fig.~\ref{fig:packaging}. Lasers and all thermal tuner contacts are wirebonded to a PCB which breaks out to control electronics. The wirebonds to the laser submount are reinforced with conductive epoxy (Circuitworks CW2400). The coupling loss between the DBR laser and PIC is 4.4 dB, quantified using the through port of the cooling SIL ring.  

\subsection*{SIL Lasers}

The optical engine consists of two atomically referenced, hybrid-integrated, SIL lasers. Both cooling and repump lasers are Modulight 780 nm distributed Bragg reflector (DBR) lasers (Modulight ML3303) with TM output polarization, bonded to the PIC. The lasers are coupled to a 2.5 million intrinsic Q add-drop ring resonator with a Sagnac loop mirror on the drop port with 10\% reflectivity. The loop mirror increases back-reflected light to the DBR laser, improving laser noise suppression and locking range \cite{su_self-injection_2023}. The phase between the DBR lasers and the ring resonator is tuned to maximize locking range and drop port output power using Pt resistive heaters. As the local temperature around the SIL ring fluctuates, the ring resonance frequency drifts. The resonance can drift outside of the SIL locking range, leading to sudden destabilization of laser frequency. To maintain stable SIL conditions, a side-of-fringe lock with feedback to laser current is engaged to maintain through-port power at a setpoint near the bottom of the SIL transmission curve \cite{liu_stabilization_2025}. The feedback effectively locks the DBR laser frequency to the ring resonance. The digital PID loop is performed using a Red Pitaya (STEMlab 125-14) running Linien \cite{Wiegand2022}, an open-source spectroscopy locking software. Once the lock is engaged, the DBR laser remains injection locked indefinitely under normal conditions, correcting for slow thermal drift, and increases robustness to crosstalk from other thermally tuned components.

\subsection*{Frequency Control}

The cooling laser is referenced to the $^{87}$Rb $5S_{1/2}\,F=2\rightarrow5P_{3/2}\,F'=3$ transition using a custom Zeeman-tunable saturated-absorption spectroscopy (SAS) setup (top of Stage 2 in Fig.~\ref{fig:architecture}). Light from an on-chip monitor tap is phase-modulated at 9.5 MHz by a fiber EOM (Thorlabs LNY7810A). The beam passes through a half-wave plate, polarizing beam splitter (PBS), and quarter-wave plate into a 7.5\,cm long $^{87}$Rb vapor cell (Thorlabs GC19075-RB87, 25\,$^\circ$C), and then is retroreflected to form counter-propagating pump and probe beams. The returning probe is separated at the PBS and detected on an avalanche photodetector (Thorlabs APD430A). A Red Pitaya (STEMlab 125-14) running Linien~\cite{Wiegand2022} generates the modulation tone, demodulates the photodiode signal to produce a dispersive error signal at each hyperfine transition, and applies PID feedback to the cooling laser's SIL ring heater. To lock the laser red-detuned from the cooling transition, which lies outside the SAS error-signal capture range, a solenoid wound around the vapor cell applies an axial magnetic field that Zeeman-shifts the SAS lock point~\cite{ho_seong_lee_frequency_1996} (Fig.~\ref{fig:control}a). The shift is calibrated by beating the locked cooling laser with an external reference laser, SAS-locked to the zero-field $F=2\rightarrow F'=3$ transition, and recording the beat frequency with a frequency counter (Keysight 53220A) as a function of coil current. The detuning is linear in current at 14.0\,MHz\,A$^{-1}$ (Fig.~\ref{fig:control}b); the MOT is operated at 1.015 A, giving a detuning of 14.2 MHz.

The repump laser frequency is controlled through a 6.58 GHz optical phase locked loop (OPLL) with the cooling laser. The repump SIL laser is photo-mixed with the SAS stabilized SIL cooling laser using an on-chip directional coupler and the beat-note is detected using a high-speed photodetector (Thorlabs DXM25CF). The high frequency beat-note is electronically mixed down to $\sim$150 MHz using a 6.43 GHz voltage-controlled oscillator (VCO) (Analog Devices HMC586LC4B). The signal is amplified and fed into a phase-frequency discriminator (RF Bay PDF-100), which produces an error signal proportional to the phase and frequency deviation of the beat-note relative to a 150 MHz low-noise VCO (Minicircuits POS200+). The error signal is filtered by a Moku:Go (Liquid Instruments) PID controller, which feeds back to the repump laser's SIL ring heater, stabilizing the repump frequency relative to the cooling laser. The repump frequency can be tuned continuously by modulating the tuning voltage applied to the reference 6.43 GHz VCO. 

\subsection*{Frequency stability measurement}

The absolute frequency stability of the SAS-locked cooling laser is measured against a self-referenced fiber frequency comb (Vescent FFC-100) whose repetition rate is phase-locked to a 1\,Hz-linewidth ULE-cavity-stabilized laser (Stable Laser Systems) / Rb-disciplined, GPS-referenced oscillator (SRS-FS725). The comb output is frequency-doubled to 780\,nm and combined with light from a cooling-laser monitor port; the heterodyne beat-note is detected on a photodiode and recorded with a frequency counter (Keysight 53220A) for 65 min (Fig.~\ref{fig:control}d). The free-running frequency trace in Fig.~\ref{fig:control}d and the repump trace in Fig.~\ref{fig:control}e are recorded with a Fizeau wavemeter (Moglabs FZW600-APC, 1-10 MHz resolution).

\subsection*{Intensity Control}

To control and stabilize the cooling laser intensity during MOT formation, a down-stream ring intensity modulator is thermally actuated to attenuate the optical signal before the gratings. The output of a higher-order diffraction from the gratings is collected by a free-space amplified photodetector (Thorlabs PDA100A2) and is used to detect intensity fluctuations. The output of the transimpedance amplifier (TIA) is acquired by a Red Pitaya running Linien and a digital PID loop forms an intensity servo controlling the attenuator’s ring heater, suppressing intensity deviations from the setpoint. The intensity servo suppresses relative intensity noise (RIN) up to the loop bandwidth of $\sim$2 kHz, shown in Fig.~\ref{fig:control}g. The servo bandwidth is limited by the thermal tuner frequency response, estimated using transient response in Extended Data Fig.~\ref{fig:thermal_tuning}c. 

\subsection*{Beam delivery}
\textbf{Beam profiling.} The emitted beams were imaged with a camera on a wax-paper screen parallel to the chip surface; the image scale (29~$\mu$m/pixel) was calibrated with a 1-mm graph-paper grid placed in the same plane. The screen was scanned in height $z$ above the chip in 0.25-mm steps. The power balance among B1--B3 and the diffraction-order ratios were obtained from a single unsaturated image taken above the beam intersection, by summing background-subtracted counts in a box around each of the six spots (the $m=1$ and $m=2$ orders of the three emitters). Beam widths are $1/e^2$ full widths of line profiles through each beam center, taken along the longitudinal ($\parallel$) and lateral ($\perp$) axes at heights above the beam intersection. Because each beam strikes the screen at 52.6$^{\circ}$ from normal incidence, its longitudinal footprint is elongated, and the measured width is multiplied by $\cos 52.6^{\circ}$ to obtain $w_{\parallel}$ in the beam frame; $w_{\perp}$ requires no correction. The emission angle and intersection height were determined by tracking each beam's center through the $z$-scan and fitting its trajectory with a straight line in three dimensions.

\textbf{Beam uniformity.} Uniformity is quantified as the r.m.s. intensity variation, $\sigma_I/\langle I\rangle$, within the central 20\% and 80\% of each beam's mode area. The mode area is bounded by the contour at 50\% of the beam's peak intensity, and the central regions are concentric, area-scaled copies of it. The grain of the paper screen adds a spurious variation larger than that of the beams themselves. To remove it, we use images from neighboring heights: because each beam walks off laterally as $z$ changes, images aligned on the beam (shift $\approx$0.3~mm, about six paper-grain radii) no longer share the paper texture, and the normalized covariance of the aligned images gives the intensity variance of the beam alone. Each estimate uses three adjacent heights with all pairs averaged; the variance is then averaged over a 0.75 mm range of heights above the beam intersection for each beam, and the largest value among B1--B3 is reported. Beam structure finer than the $\approx$60$~\mu$m smoothing scale is not resolved.

\textbf{Slab expansion.} Before each emitter, the 900-nm routing waveguide is tapered to 400~nm and then expanded into the slab region that feeds the grating. Propagation from the end of the taper into the slab was simulated with the Tidy3D finite-difference time-domain (FDTD) solver over a 20-$\mu$m-long region. The transverse intensity profile was extracted at a series of distances along the slab; its full width at half maximum (FWHM) grows approximately linearly with propagation distance at a rate of $\approx$0.7, corresponding to a full divergence angle of $\approx$39$^{\circ}$ in the slab.

\subsection*{Polarization and magnetic field}

The grating emitters produce vertically polarized output beams, which are converted to circular polarization to drive the $\sigma^\pm$ transitions required for the 3D-MOT. A quarter-wave plate (QWP; Tower Optical Z-10-A-.250-N-780) was mounted above each grating emitter in a machined metal holder anodized black to suppress stray reflections. Each QWP was rotated while monitoring the transmitted beam with a Schäfter+Kirchhoff SK010PA-VIS polarimeter and fixed in place with UV-curable epoxy at the angle that maximizes the desired circular-polarization component. The holder also serves as a spatial aperture that blocked the nearly vertically emitted higher diffraction orders. After UV curing, the beam polarizations were remeasured. The degree of circular polarization is defined as $\mathrm{DOCP}=|S_3|/S_0$, where $S_3$ is the circular Stokes component and $S_0$ is the total optical intensity. The measured DOCP values were 90.2\% for B1 (RCP), 96.4\% for B2 (LCP), and 86.6\% for B3 (RCP).

The quadrupole field for the 3D-MOT is generated using a six-coil configuration based on the geometry of \cite{comptonMagneticFieldCoils2014}. Each coil consists of a stack of three two-layer inductors (Vishay IWTX47R0EBEB240JR1) wired in series. The three upper coils and three lower coils form independently driven sets, with the current direction of each coil chosen to produce the required quadrupole-field orientation relative to the corresponding beam polarization. Viewed from above, the upper-layer currents are CW for B1, CCW for B2, and CW for B3, while the lower-layer currents are reversed. The upper and lower coil sets are driven independently by separate channels of an Agilent E3648A power supply, with a nominal current of 5~A applied to each set. Their measured resistances are 0.9~$\Omega$ and 1.2~$\Omega$, respectively, corresponding to a total electrical power consumption of $P = I^2(R_{\rm top}+R_{\rm bottom}) \approx 52.5$~W. Beam B2 is aligned with the quadrupole-field axis. The measured magnetic-field gradient is 21.5~G/cm along B2 and 10.2~G/cm along B1 (Extended Data Fig.~\ref{fig:bfield_grad}).

\subsection*{System control and sequencing}
% Short section about the lock loop infrastructure
All lock loops and actuators are run from FPGA-based controllers and a host computer. Each SIL laser is held in the SIL state by its own Red Pitaya (STEMlab 125-14) running Linien~\cite{Wiegand2022}, which applies side-of-fringe feedback to the DBR current. A third Red Pitaya runs the Zeeman-SAS lock of the cooling laser, with feedback to the cooling SIL ring heater, and a fourth runs the intensity servo, with feedback to the ring-modulator heater. The repump offset lock uses a Moku:Go (Liquid Instruments)~\cite{mokugo2026} as the loop filter, acting on the phase-frequency-detector error signal and driving the repump SIL ring heater. 

Locks are engaged in sequence: (i) both SIL locks, (ii) the cooling-laser SAS lock at the Zeeman-set detuning, (iii) the repump offset lock, and (iv) the intensity servo at the chosen MOT beam power. Once engaged, the MOT runs continuously without further intervention. For time-of-flight and absorption-imaging measurements, a pulse generator (PulseBox~\cite{hosak2018Arbitrary}) synchronizes the experimental cycle. The MOT coils are switched off together with the cooling light. Because the usable modulator extinction is limited to 2.2\,dB (Discussion), the cooling light is extinguished by switching off the cooling DBR current with a custom current-shutoff circuit for 100\,ms, while the repump remains on. The probe pulse and camera exposure are then triggered after a variable delay $t_\mathrm{TOF}$. 

A programmable Arduino-based pulse generator based on \cite{hosak2018Arbitrary} controls the timing of the magnetic field shutoff and laser shutoff, the triggering of the cameras, scopes, and photodetectors, and an AOM which shuttered the on-resonance probe beam used for absorption imaging.

\subsection*{Power Budget}

The size and electrical power consumption of the principal PICMOT system components are summarized in  Extended Data Table~\ref{tab:swapc}. The estimated total power consumption is 89.25~W, dominated by the magnetic field coils (52.5~W) and the miniMOT physics package (33~W). The six-coil quadrupole magnetic field configuration accounts for the largest individual power contribution, while the miniMOT power consumption is estimated from its specified operating current of 300~mA at 110~V. The two semiconductor DBR lasers operate at injection currents of approximately 240~mA, with an estimated combined electrical power consumption of 1~W for the laser current drivers. The ring-resonator heater drivers consume approximately 0.95~W, while the thermoelectric cooler (TEC), used to stabilize the packaged photonic light engine at 25~$^\circ$C, consumes approximately 0.2~W. The Zeeman-tunable saturated-absorption spectroscopy (SAS) reference coils contribute an additional 1.6~W.

The magnetic field coils and vacuum physics package account for approximately 96\% of the estimated total system power consumption, while the remaining laser and photonic control components contribute approximately 4\%. These estimates reflect the present laboratory implementation, in which several electronic control and reference subsystems remain external to the packaged PIC. Further reductions in system size and power consumption could be achieved through miniaturization of the magnetic field coils, integration of the laser control electronics, and optimization of the thermal management architecture.

\subsection*{MOT characterization}

\textit{Atom number.} Measurement of MOT atom number is performed using two methods, fluorescence detection and absorption imaging, outlined in detail in previous work \cite{isichenko2023Photonica}. For fluorescence detection, the MOT fluorescence is focused and collected on a free-space amplified photodetector (Thorlabs PDA36A2) during the MOT formation stage. Simultaneously, the cooling laser detuning is measured relative to an external reference laser, SAS locked to the cooling transition, by measuring the beat frequency of the two lasers with a frequency counter (Keysight 53220A). The peak-peak power change during loading and cooling laser detuning can be converted to atom number using 
\begin{equation}
    N_0 = \frac{V_0}{\eta\, R(\lambda)\, G} \frac{4\pi}{\Omega} \frac{1}{R_{\mathrm{scat}}\, (hc/\lambda)} ,
    \label{eq:N0_from_voltage}
\end{equation}
where $V_0$ is the peak-peak voltage output from photodetector, $\eta$ is collection efficiency, R($\lambda$) is responsivity of the photodetector, G is the gain of the photodetector TIA, $\Omega$ is the solid angle for the collection optics, $R_{\mathrm{scat}}$ is the scattering rate, $h$ is Planck's constant, $c$ is the speed of light, and $\lambda$ is the wavelength of the fluorescence.

The second method used to determine the MOT atom number is absorption imaging. An external probe laser, stabilized to the cooling transition using saturated absorption spectroscopy (SAS) at a detuning of $\Delta_{\mathrm{probe}} \approx 0$, illuminates the atom cloud 0.9 ms after the magnetic field and cooling laser are switched off. The transmitted probe light is recorded using a CMOS camera (Thorlabs CS165MU). Probe pulses are generated using an acousto-optic modulator (AOM, Aerodiode 780AOM-2) controlled by an RF switch (Mini-Circuits ZASWA-2-50DRA+). The probe pulse duration is 300 $\mu$s, with an optical power of 14 $\mu$W stabilized using an intensity servo. The camera exposure time is set to 200 $\mu$s.

A second image is acquired using an identical probe pulse 70 ms after MOT release, when the atom cloud has dissipated, to provide a reference image without atoms. The two images are used to calculate the spatial optical density (OD) distribution using the Beer–Lambert law. Assuming a Gaussian spatial distribution of the atomic density, the root-mean-square (r.m.s.) cloud widths, $\sigma_y$ and $\sigma_z$, are extracted from a two-dimensional Gaussian fit to the OD distribution. The atom number is then calculated by integrating the OD distribution over the imaging plane:

\begin{equation}
    N_0 = \frac{OD_{max}}{\sigma}2\pi\sigma _y \sigma _z ,
    \label{eq:N0_from_absorb_img}
\end{equation}
where $\mathrm{OD}_{\max}$ is the peak optical density and $\sigma$ is the effective atomic absorption cross-section at the probe detuning.

\textit{Temperature.} The MOT temperature is measured with the time-of-flight imaging technique. For this measurement, absorption images, as described in the atom number measurement, are taken at increasing flight times after the trap is shut-off. We record flight times from 0.5 ms to 2.0 ms. From the images, the r.m.s. widths of the atom cloud are measured for the two axes imaged as the MOT expands. With ballistic expansion of the MOT, the temperature of the cloud can be determined by the increase of the atom cloud's r.m.s. width as a function of flight time,

\begin{equation}
\begin{split}
    \sigma_y ^2 = \sigma^2_{y,i} + \frac{k_B T_y}{m}t_{\mathrm{TOF}}^2 \\
    \sigma_z ^2 = \sigma^2_{z,i} + \frac{k_B T_z}{m}t_{\mathrm{TOF}}^2 
\end{split}
    \label{eq:N0_from_absorb_img}
\end{equation}
where $\sigma_{y,i}$ and $\sigma_{z,i}$ are the initial cloud widths, $k_B$ is the Boltzmann constant, and $m$ is the mass of a single $^{87}$Rb atom. 

\textit{Statistics.} Temperatures $T_y$ and $T_z$ are obtained from weighted linear fits of $\sigma^2$ versus $t_{\mathrm{TOF}}^2$ (Fig.~\ref{fig:MOT}c; nine flight times, five images averaged per flight time); the quoted uncertainties are the $1\sigma$ standard errors of the fitted slopes. The fluorescence atom number $N = (1.12 \pm 0.16)\times10^6$ uncertainty is propagated from the uncertainties in collection efficiency, solid angle, photodetector responsivity, and detuning. Error bars in Fig.~\ref{fig:MOT}e are the propagated uncertainty of each measurement.

\subsection*{Data and code availability}
The data and code generated in this study are available from the corresponding authors upon reasonable request.

%%===========================================================================================%%
%% If you are submitting to one of the Nature Portfolio journals, using the eJP submission   %%
%% system, please include the references within the manuscript file itself. You may do this  %%
%% by copying the reference list from your .bbl file, paste it into the main manuscript .tex %%
%% file, and delete the associated \verb+\bibliography+ commands.                            %%
%%===========================================================================================%%
\bibliography{sn-bibliography}

\section*{Acknowledgments}
We thank Simon Soller for help with test-structure measurements, Josh Hansen for development of the current-shutoff circuit, and Steven Carpenter for help with PulseBox development. We thank Modulight for providing initial laser samples, and Jen Choy at the University of Wisconsin and the NASA Quantum Pathways team for useful discussions. We thank Demis John for assistance with ASML stepper programming at the UCSB Nanofabrication Facility, and Makoto Endo, Kirsten Okamoto, and Marcell Peña for their contributions to the PICMOT mechanical assembly and the hybrid-integration packaging station. We thank Nicholas Hutzler at Caltech for his comments on the manuscript. A portion of this work was performed in the UCSB Nanofabrication Facility, an open-access laboratory.

\section*{Funding}

This research was supported in part by the NASA Quantum Pathways Institute (80NSSC23K1343), the NSF Quantum Leap Challenge Institute Q-SEnSE (OMA-2016244), NSF National Quantum Virtual Laboratory (2533041), NSF QuSeC-TAQS (2326784), and the Air Force Research Laboratory through the University of New Mexico COSMIAC Research Center (award FA9453-20-2-001, subaward 282109-873R). S.J.H. acknowledges funding from the EPSRC Centre for Doctoral Training in Quantum Engineering (EP/S023607/1). Any opinions, findings, conclusions, or recommendations expressed in this material are those of the authors and do not necessarily reflect the views of the U.S. Government or the National Aeronautics and Space Administration. The U.S. Government is authorized to reproduce and distribute reprints for Government purposes, notwithstanding any copyright notation herein.

\section*{Competing interests}
D.J.B.’s work has been funded in the past by Infleqtion and he has consulted for Infleqtion and owns stock. All other authors declare no competing interests.

\section*{Author Contributions}

D.J.B. and A.I. conceived the work and directed the research. A.I. designed the PICMOT silicon nitride photonic integrated circuit. C.Z. fabricated the photonic devices, with K.D.N. performing the upper-cladding deposition and M.S. contributing to nanofabrication process development and fabrication of earlier beam-delivery PICs. A.I. and C.R. characterized the PIC components, performed hybrid integration of the self-injection-locked lasers, and conducted laser characterization and frequency-noise measurements. C.R. and A.I. developed the laser stabilization and control software. N.M. and C.L. developed the experimental timing system. A.I., C.R. and N.M. performed the cold-atom experiments, with S.J.H. contributing to atom-number and temperature characterization. A.I., C.L. and S.J.H. performed magnetic-field simulations, measurements and assembly. N.I. contributed to magnetic-field simulations and the design and assembly of the PICMOT mounting chassis. D.J.B., A.I. and C.R. wrote the manuscript, with contributions from C.Z. All authors reviewed and approved the manuscript.

\section*{Materials and Correspondence}
Correspondence to Daniel J. Blumenthal (danb@ucsb.edu) and Andrei Isichenko (andrei.isichenko@gmail.com). 

\newpage
\captionsetup[table]{name=Extended Data Table}
\section*{Tables}

\begin{table}[h]
\caption{Critical-dimension measurements of the fabricated PICMOT device. The design target dimensions are compared with dimensions measured by SEM.}
\begin{tabular*}{\textwidth}{@{\extracolsep\fill}lcccccc}
\toprule%
\textbf{Feature} & \textbf{Design (nm)} & \textbf{Measured (nm)} & 
\textbf{$|\Delta|$ (nm)} \\
\hline
Beam expander 1 & 400  & 427  & 27 \\
Beam expander 2 & 400  & 404  & 4  \\
Beam expander 3 & 400  & 419  & 19 \\
Gap 1           & 1000 & 967  & 33 \\
Gap 2           & 1000 & 969  & 31 \\
Input taper 1   & 250  & 258  & 8  \\
Gap 3           & 800  & 784  & 16 \\
Input taper 2   & 250  & 251  & 1  \\
Gap 4           & 800  & 810  & 10 \\
Gap 5           & 1500 & 1464 & 36 \\
\botrule
\end{tabular*}
\label{SEMs_CD}
\end{table}

\begin{table}[h]
\caption{Size and power consumption estimates for main components in the PICMOT system.}
\begin{tabular*}{\textwidth}{@{\extracolsep\fill}lcccccc}
\toprule%
\textbf{Component} & \textbf{Power Consumption (W)} & \textbf{Size (\textbf{$cm^3$)}} \\
\hline
MOT coils & 52.5 & 30.1  \\
Physics package (miniMOT) & 33  &  34.4\\
Zeeman SAS coils & 1.6  & 450  \\
Laser current drivers (2 used in system) & 1 & 10,036  \\
Current driver for ring heaters & 0.95 & 7.6  \\
TEC for temperature control & 0.2 & 0.4  \\
\botrule
\end{tabular*}
\label{tab:swapc}
\end{table}

\section*{Extended Data Figures}

\setcounter{figure}{0}
\renewcommand{\thefigure}{\arabic{figure}}
\captionsetup[figure]{name=Extended Data Fig.,labelsep=colon}

\begin{figure}[h]
\centering
\includegraphics[width=0.9\textwidth]{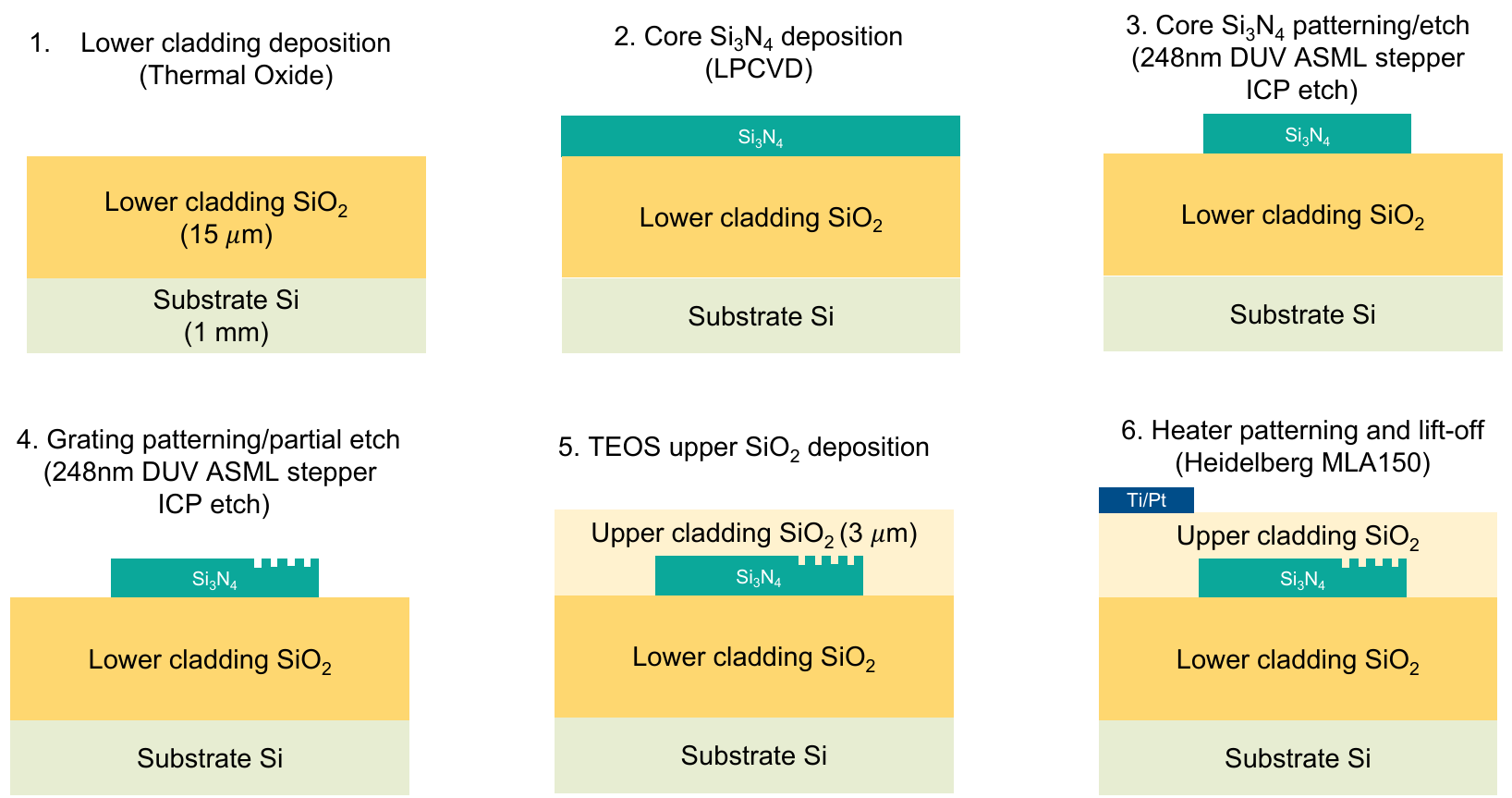}
\caption{\textbf{PICMOT Beam delivery fabrication process flow.} Step 1: lower SiO$_2$ cladding deposition. Step 2: Silicon nitride core deposition. Step 3: Silicon nitride core etch. Step 4: grating layer partial etch. Step 5: Upper oxide cladding deposition. Step 6: Heater patterning, metal deposition and lift-off.}
\label{fig:Process_flow}
\end{figure}

\begin{figure}[h]
\centering
\includegraphics[width=0.9\textwidth]{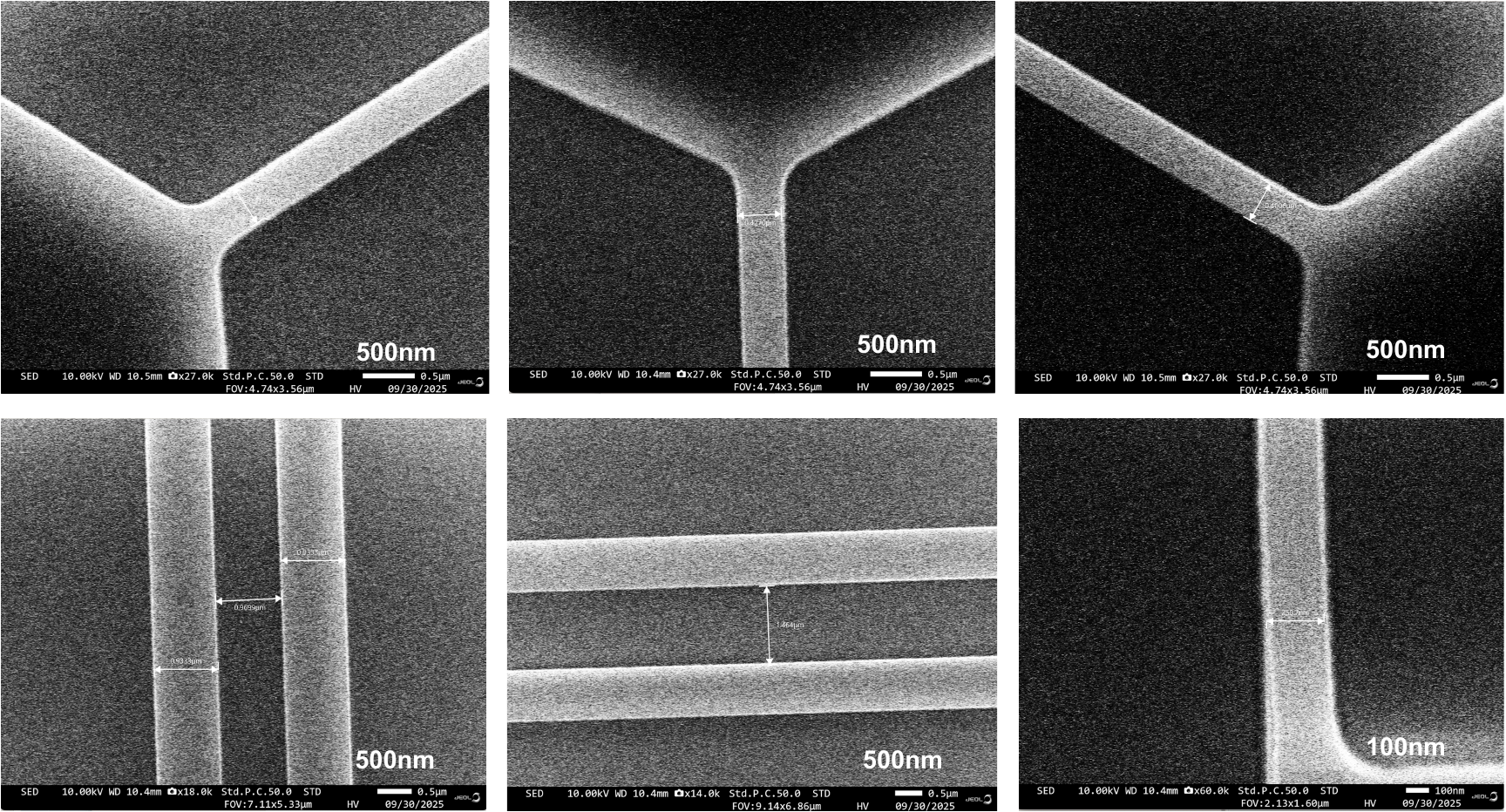}
\caption{\textbf{SEM characterization of the critical dimensions of the PICMOT subsections.} \textbf{Top row,} Input width of the three beam expander. \textbf{Bottom row,} Critical dimensions of the directional couplers with 0.8 ${\mu}$m and 1.5 ${\mu}$m gap. Input width of the 250-nm-wide input taper.}
\label{fig:SEMs}
\end{figure}

\begin{figure}[h]
\centering
\includegraphics[width=1.0\textwidth]{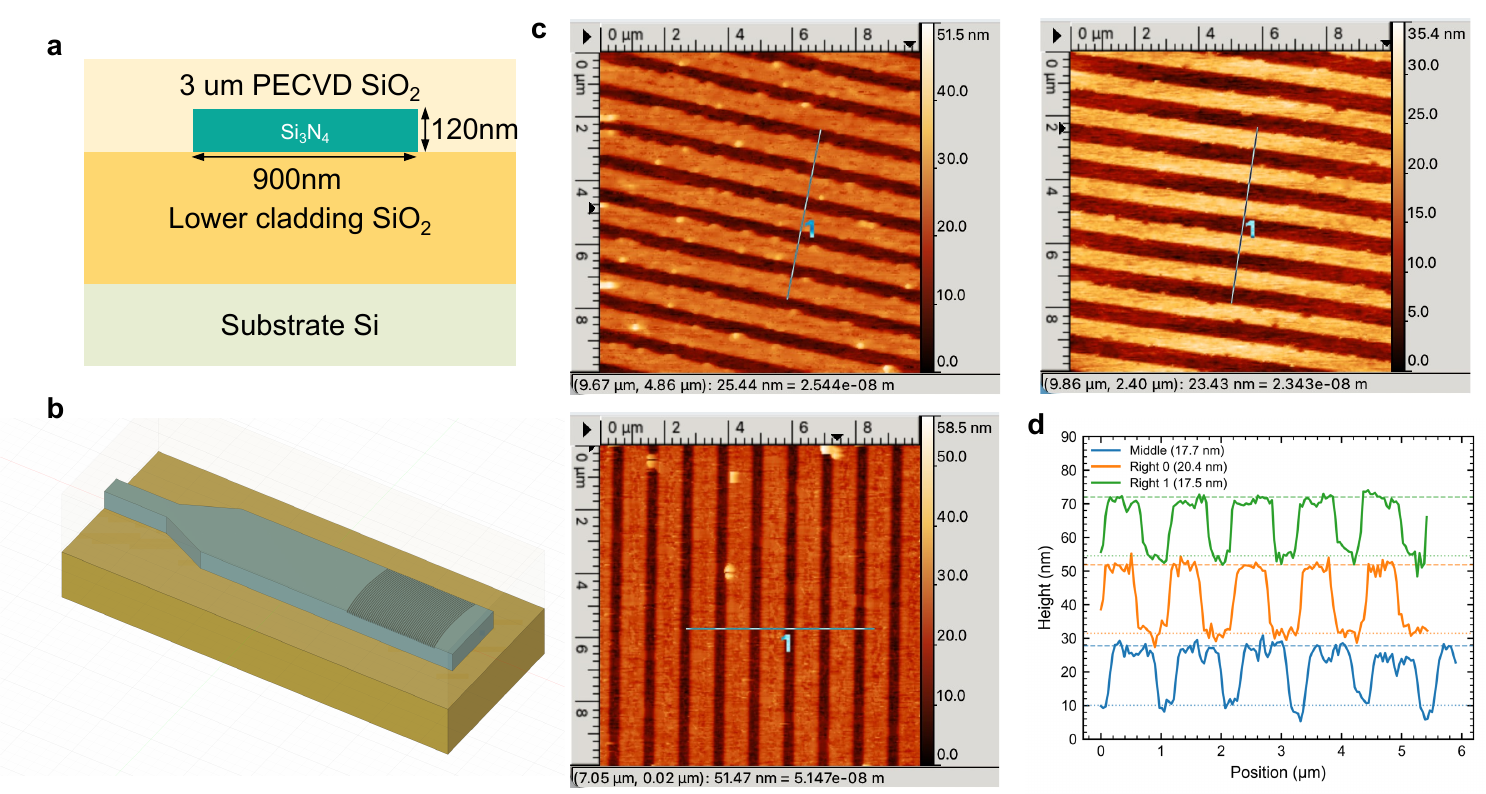}
\caption{\textbf{AFM images of grating partial etch.} \textbf{a,} Waveguide dimension for TM$_0$ operation at 780 nm.
\textbf{b,} Slab expander and grating emitter with 18.5~nm partial etch. \textbf{c,} AFM images acquired at three locations on the partially etched grating emitter. 
\textbf{d,} Height profiles measured at the three AFM locations, showing an average grating-emitter etch depth of 18.5~nm.}
\label{fig:AFMs}
\end{figure}

\begin{figure}[h]
\centering
\includegraphics[width=0.8\textwidth]{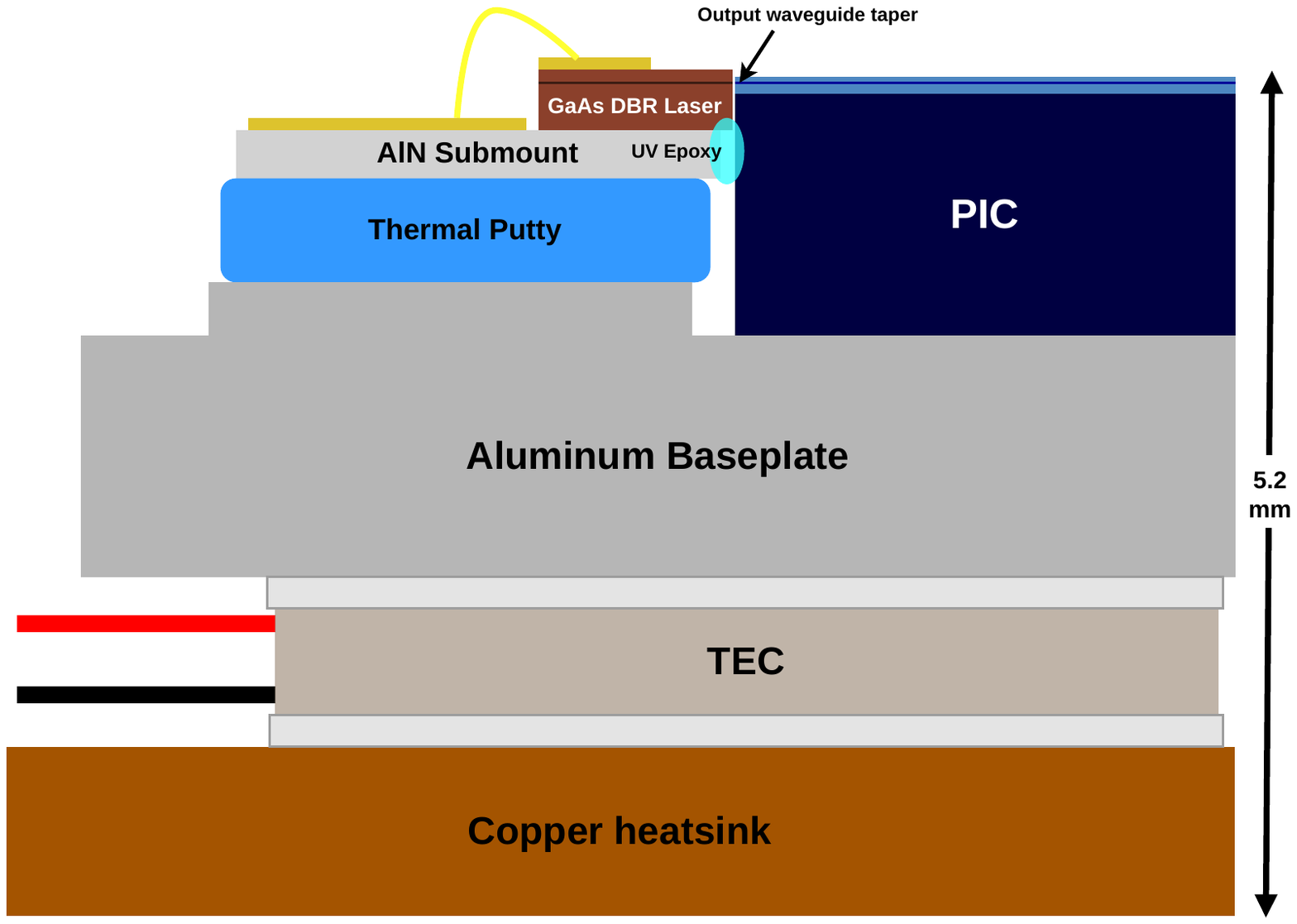}
\caption{\textbf{Hybrid integrated laser packaging.} Side profile of packaging stack used in laser packaging. TEC: thermoelectric cooler, DBR: distributed Bragg reflector.}
\label{fig:packaging}
\end{figure}

\begin{figure}[h]
\centering
\includegraphics[width=0.9\textwidth]{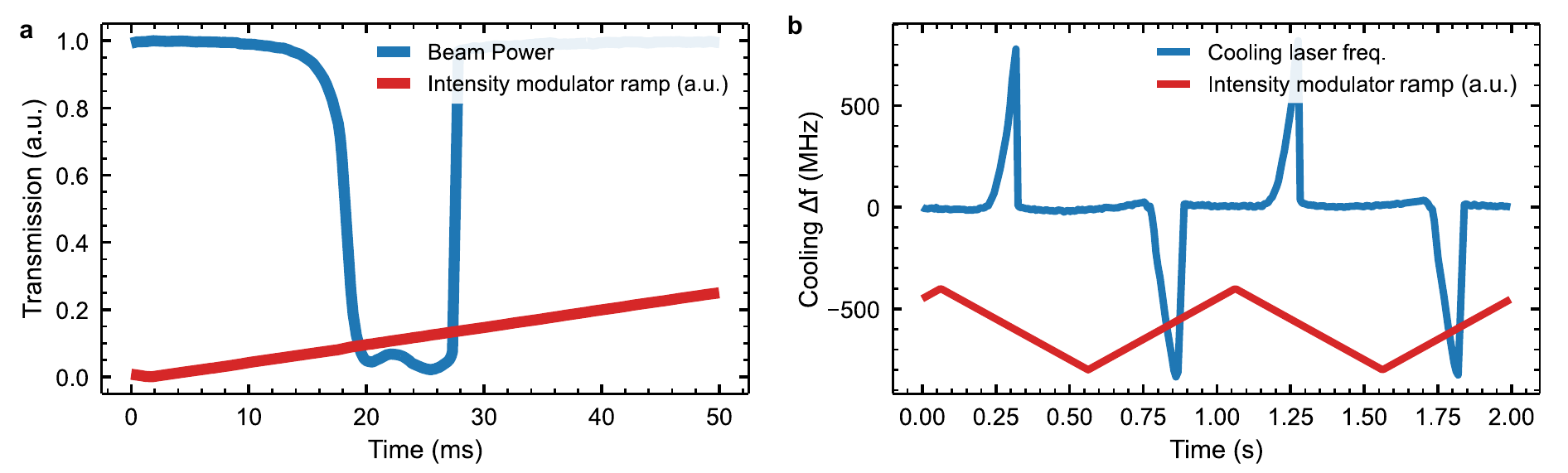}
\caption{\textbf{Intensity modulator feedback.} \textbf{a,} Free-space beam power measured during linear ramp of intensity modulator across the cooling SIL laser frequency (repump laser is off). The transmission spectrum has characteristic self-injection lock shape. \textbf{b,} Cooling laser frequency monitored with a wavemeter during sequential intensity modulator ramps. Laser frequency is pulled in and follows intensity modulator resonance, increasing or decreasing frequency depending on ramp direction. }
\label{fig:shutter_feedback}
\end{figure}

\begin{figure}[h]
\centering
\includegraphics[width=0.9\textwidth]{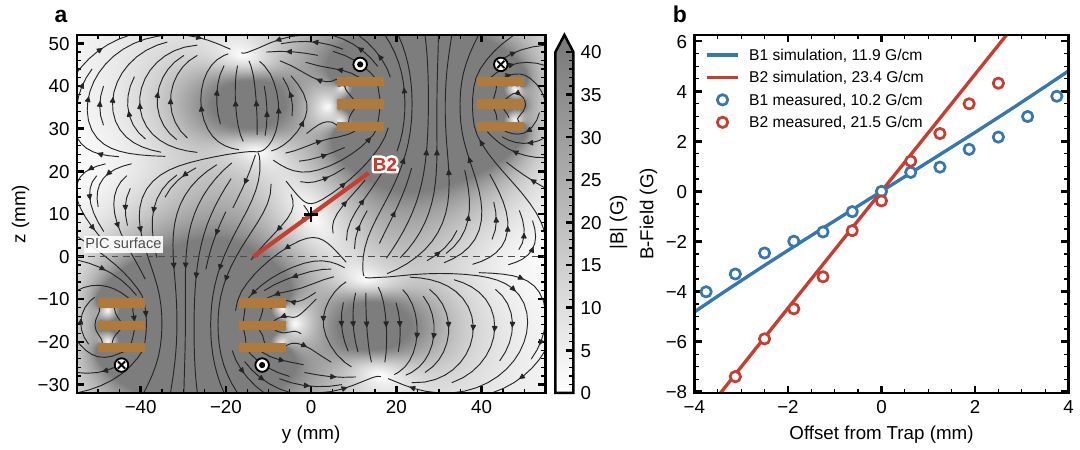}
\caption{\textbf{Magnetic field gradient.}
\textbf{a,} Simulated quadrupole magnetic field generated by the six-coil configuration at a current of 5 A per coil set (Magpylib 5.1.1). Each coil is modeled with 6 layers of 10 turns at its corresponding position in the three-coil stack.
\textbf{b,} Simulated and measured magnetic field profiles along beam paths B1 and B2 through the MOT trapping region. Measurements are performed using a scalar magnetometer. The simulated field gradients are 11.9 and 23.4 G cm$^{-1}$ along B1 and B2, respectively, compared with measured gradients of 10.2 and 21.5 G cm$^{-1}$.}
\label{fig:bfield_grad}
\end{figure}

\begin{figure}[h]
\centering
\includegraphics[width=0.9\textwidth]{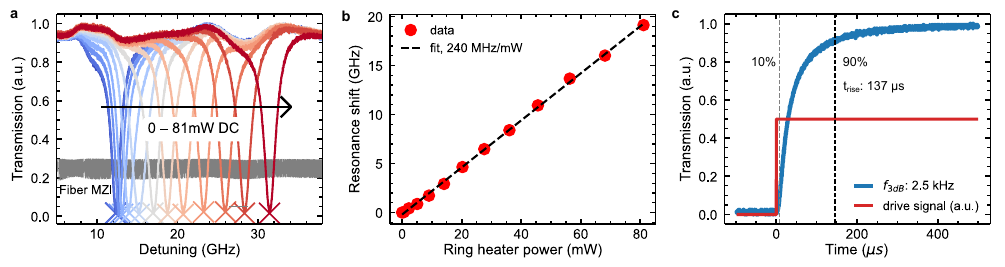}
\caption{\textbf{Thermal ring tuner response.}
\textbf{a,} Transmission spectrum through an add-drop ring resonator test structure, with similar geometry to the on-chip self-injection lock rings, at increasing heater power. The free-spectral range (FSR) of the fiber MZI (grey) is 11.3 MHz.   
\textbf{b,} Resonance detuning from sweep in \textbf{a} as a function of ring heater power. Tuning strength of 240 MHz mW$^{-1}$ is measured from a linear fit.
\textbf{c,} Transient response of thermally tuned ring test structure with a step function applied. A 3dB bandwidth of 2.5 kHz is extracted from the transmission using the 10\% to 90\% rise time, assuming a single-pole response ($f_{3dB}=0.35~t^{-1}_{rise}$).
}
\label{fig:thermal_tuning}
\end{figure}

\begin{figure}[h]
\centering
\includegraphics[width=0.9\textwidth]{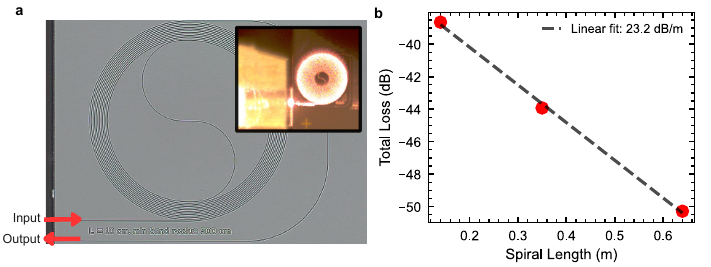}% loss spiral yielding 23 dB/m. 
\caption{\textbf{Optical propagation loss measurement using spiral test structures.}
\textbf{a,} Optical micrograph of a spiral waveguide test structure, with input and output facets indicated by red arrows. Inset shows the device under test during optical characterization. 
\textbf{b,} Measured optical transmission loss as a function of spiral waveguide length. A linear fit to the measured data yields a waveguide propagation loss of 23.2~dB m$^{-1}$.
}
\label{fig:PIC_measurements}
\end{figure}

\end{document}